\newif\ifrevisioncolors
\revisioncolorsfalse
\newcommand{\arxivlayout}{}

\AtBeginDocument{%
  \setlength{\bibsep}{3pt plus 1pt minus 1pt}%
  \fancyhead{}%
  \author{%
    \begin{tabular}[t]{l}
      \textbf{Ziv Aharoni, Henry D. Pfister}\\
      Department of Electrical and Computer Engineering\\
      Duke University, USA\\
      \texttt{\{ziv.aharoni,henry.pfister\}@duke.edu}
    \end{tabular}}%
}
\documentclass{article}
\usepackage{iclr2027_conference,times}
\usepackage{amsmath,amssymb,amsthm,booktabs}
\usepackage[normalem]{ulem}
\usepackage{graphicx}
\usepackage{float}
\usepackage{placeins}
\usepackage{wrapfig}
\usepackage{needspace}
\usepackage{capt-of}
\usepackage{algorithm}
\usepackage{algpseudocode}
\usepackage{tikz}
\usetikzlibrary{arrows.meta,positioning,fit,shapes.geometric}
\usepackage{pgfplots}
\usepgfplotslibrary{groupplots,fillbetween}
\pgfplotsset{compat=1.18}
\usepackage{hyperref}
\hypersetup{hidelinks}
\usepackage{url}

\makeatletter
\@ifundefined{ifrevisioncolors}{%
  \newif\ifrevisioncolors
  \revisioncolorstrue
}{}
\makeatother
\newcommand{\revisiontext}[2]{%
  \ifrevisioncolors{\color{#1}#2}\else#2\fi}
\AtBeginDocument{%
  \ifrevisioncolors\else
    \colorlet{draftcolor}{black}%
    \colorlet{revisioncolor}{black}%
    \colorlet{revision2color}{black}%
    \renewcommand{\henrytask}[2]{#2}%
  \fi}

\definecolor{draftcolor}{RGB}{150,55,145}
\definecolor{revisioncolor}{RGB}{0,82,155}

\newcommand{\revision}[1]{\revisiontext{revisioncolor}{#1}}
\definecolor{revision2color}{RGB}{0,110,95}
\let\revisionone\revision
\makeatletter
\DeclareRobustCommand{\revision}{\@ifnextchar2{\revisionsecond}{\revisionone}}
\def\revisionsecond2#1{\revisiontext{revision2color}{#1}}
\makeatother

\DeclareRobustCommand{\henrytask}[2]{{\color{revisioncolor}\textsuperscript{\normalfont\scriptsize[T#1]}#2}}

\newcommand{\draftbibkey}[1]{%
  \expandafter\def\csname bibreviewcolor@#1\endcsname{draftcolor}}
\newcommand{\revisionbibkey}[1]{%
  \expandafter\def\csname bibreviewcolor@#1\endcsname{revisioncolor}}
\AddToHook{env/thebibliography/begin}{%
  \let\originalbibitem\bibitem
  \RenewDocumentCommand{\bibitem}{o m}{%
    \par
    \ifcsname bibreviewcolor@#2\endcsname
      \edef\currentbibreviewcolor{\csname bibreviewcolor@#2\endcsname}%
      \color{\currentbibreviewcolor}%
    \else
      \color{black}%
    \fi
    \IfNoValueTF{#1}{\originalbibitem{#2}}{\originalbibitem[#1]{#2}}}}

\title{Beam Search for Error-Correcting Codes\\via Conditional Diffusion}

\DeclareRobustCommand{\mbsresult}[1]{MBS\ifdefined\arxivlayout\ ($L=#1$)\fi}

\colorlet{draftcolor}{black}
\colorlet{revisioncolor}{black}
\renewcommand{\henrytask}[2]{#2}

\newcommand{\preprintupdatedresults}{}

\iclrfinalcopy
\author{%
  \begin{minipage}[t]{0.48\textwidth}
    \centering
    \textbf{Ziv Aharoni}\\
    Duke University
  \end{minipage}\hfill
  \begin{minipage}[t]{0.48\textwidth}
    \centering
    \textbf{Henry D. Pfister}\\
    Duke University
  \end{minipage}}
\newcommand{\papercodelink}{\url{https://github.com/zivaharoni/conditional-diffusion-ecc}}
\makeatletter
\renewcommand{\@maketitle}{%
  \vbox{\hsize\textwidth
    {\LARGE\scshape\@title\par}%
    \ifx\@author\@empty\else
      \vskip 0.15in
      \noindent\@author\par
    \fi
    \vskip 0.3in minus 0.1in}}
\makeatother
\hypersetup{pdftitle={Beam Search for Error-Correcting Codes via Conditional Diffusion},
  pdfauthor={Ziv Aharoni and Henry D. Pfister}}

\begin{document}
\raggedbottom
\maketitle
\begin{abstract}
We introduce a diffusion framework for neural decoding of error-correcting codes (ECCs).
We interpret decoding as a conditional diffusion process that progressively predicts codeword bits while treating the physical channel observation as fixed side information.
At each stage, a neural model estimates bitwise posteriors conditioned on the physical observation and the current noisy \revision2{state} of the codeword.
\henrytask{8}{We focus particularly on masked diffusion, for which the reverse process naturally induces a Maxwell beam search \revision2{(MBS)} procedure inspired by the Maxwell construction of Measson et al.}
Rather than committing to a single uncertain decision, the decoder maintains multiple candidate trajectories, removes parity-infeasible candidates, and ranks the remaining candidates using accumulated reverse-transition log-probabilities.
We evaluate the framework across multiple code families on binary phase-shift keying (BPSK) over additive white Gaussian noise (AWGN).
Experiments show competitive decoding performance across Polar, low-density parity-check (LDPC), and Bose--Chaudhuri--Hocquenghem (BCH) codes.
The \revision2{MBS} decoder consistently improves upon the corresponding single-trajectory diffusion decoder and, in \revision2{all} evaluated settings, improves upon available neural decoding results.
\end{abstract}

\providecommand{\papercodelink}{\nolinkurl{https://anonymous.4open.science/r/conditional-diffusion-ecc-B6FA}}
\begingroup
\renewcommand{\thefootnote}{}
\footnotetext{Code in \papercodelink}
\endgroup

\section{Introduction}
\looseness=-1
Reliable communication requires recovering transmitted information from noisy channel observations with low error and acceptable decoding latency.
For a uniform prior over codewords, the optimal decision under frame-error loss is maximum-likelihood (ML) decoding.
It is well-known that finding an ML codeword for a general linear code is NP-hard \citep{berlekamp1978inherent}.
To avoid the cost of exact ML computation, practical decoders exploit the structure of specific code families.
For low-density parity-check (LDPC) codes, belief propagation (BP) passes messages on a sparse factor graph and the min-sum approximation simplifies the message updates \citep{kschischangFactorGraphsSumProduct2001,fossorier1999reduced}.
This complexity reduction comes with tradeoffs: BP loses accuracy on graphs with cycles, and min-sum introduces additional \revision2{approximation error}~\citep{yedidiaConstructingFreeenergyApproximations2005,fossorier1999reduced}.
\revision2{One way to mitigate the limitations of BP on graphs with cycles is to train a neural network to learn message updates from data.}
Indeed, on selected codes, learned decoders improve classical message-passing performance \citep{nachmaniLearningDecodeLinear2016,nachmaniAutoregressiveBeliefPropagation2021,parkCrossMPTCrossAttentionMessagePassing2025}.
\par

Despite the success of BP-based neural decoders for sparse graph codes, polar codes are commonly decoded using \revision2{a} successive cancellation (SC) \revision2{decoder}, which decides codeword bits sequentially \citep{arikanChannelPolarization2009}.
An incorrect early decision cannot be revisited, which yields suboptimal performance.
Successive-cancellation list (SCL) decoding \citep{talListDecodingPolar2015} addresses this limitation by branching at bit decisions, retaining multiple candidate paths, and selecting a codeword only after all bits have been processed using a metric that corresponds to their likelihood.
This guessing mechanism turns the suboptimal SC decoder into a competitive decoder.
\revision2{For polar codes, SCL decoding outperforms neural decoding methods, particularly as the block length increases.}
Its main drawback is the sequential dependence inherited from SC, which limits parallelism and can increase latency\footnote{List paths can be processed in parallel, but the bit decisions along each SC path depend on earlier decisions. }.
The benefits of list decoding for SC decoding for polar codes thus lead to the central question of this paper: Can we devise a principled guessing mechanism that improves neural decoders by spending additional computation in parallel across candidates, rather than in additional sequential steps?
\par

\begin{figure}[t]
  \centering
  \begin{minipage}[t]{0.48\linewidth}
    \centering
    \vspace{0pt}
    \begin{minipage}[c][0.94\linewidth][c]{\linewidth}
    \centering
\begin{tikzpicture}
\begin{axis}[
  xbar,bar width=4.3pt,bar shift=0pt,
  width=0.80\linewidth,height=0.99\linewidth,
  xmin=0,xmax=9.8,ymin=-0.7,ymax=16.7,y dir=reverse,
  xtick={0,2,4,6,8},
  ytick={0,1,2,3,4,5,6,7,8,9,10,11,12,13,14,15,16},
  yticklabels={{BP-5},{BP-10},{BP-20},{BP-50},{SCL-1},{SCL-4},{SCL-8},{SCL-16},{SCL-32},{SCL-64},{ECCT},{CrossMPT},{DDECC},{SB-ECC},{\textbf{Conditional Diffusion}},{\textbf{\revision2{MBS}\revision2{, $L=64$}}},{OSD-3 estimate}},
  tick label style={font=\tiny},
  xlabel={$-\ln(\mathrm{BER})$},xlabel style={font=\scriptsize},
  xmajorgrids,grid style={gray!20},
  axis x line*=bottom,axis y line*=left,
  tick align=outside,
  nodes near coords,nodes near coords align={horizontal},
  every node near coord/.append style={font=\tiny,fill=white,inner sep=1pt,/pgf/number format/fixed,/pgf/number format/precision=2},
]
\addplot[fill=blue!65,draw=none] coordinates {(3.5265055,0)};
\addplot[fill=blue!65,draw=none] coordinates {(3.8011284,1)};
\addplot[fill=blue!65,draw=none] coordinates {(3.9959645,2)};
\addplot[fill=blue!65,draw=none] coordinates {(4.2920658,3)};
\addplot[fill=gray!65,draw=none] coordinates {(3.8083000,4)};
\addplot[fill=gray!65,draw=none] coordinates {(5.8792000,5)};
\addplot[fill=gray!65,draw=none] coordinates {(6.8695000,6)};
\addplot[fill=gray!65,draw=none] coordinates {(7.5590000,7)};
\addplot[fill=gray!65,draw=none] coordinates {(7.8346000,8)};
\addplot[fill=gray!65,draw=none] coordinates {(7.9150000,9)};
\addplot[fill=teal!65,draw=none] coordinates {(6.9900000,10)};
\addplot[fill=teal!65,draw=none] coordinates {(7.5000000,11)};
\addplot[fill=teal!65,draw=none] coordinates {(7.0300000,12)};
\addplot[fill=teal!65,draw=none] coordinates {(7.7700000,13)};
\addplot[fill=orange!75,draw=none,every node near coord/.append style={font=\tiny\bfseries,text=revisioncolor}] coordinates {(7.7028921,14)};
\addplot[fill=red!65,draw=none,every node near coord/.append style={font=\tiny\bfseries,text=revision2color}] coordinates {(7.922924046002467,15)};
\addplot[fill=black!70,draw=none] coordinates {(7.9783269,16)};
\end{axis}
\end{tikzpicture}
    \vspace{-4mm}
    \end{minipage}
\captionof{figure}{\henrytask{1}{\revision2{Comparison of $-\ln(\mathrm{BER})$} on Polar$(64,32)$ at 4 dB; higher is better. \revision2{\revision2{MBS} uses beam size 64.} Order-3 ordered-statistics decoding (OSD-3) \revision2{approximates ML decoding}. \revision2{Details are given in Appendix~\ref{app:polar64-ber-sources}.}}}
    \label{fig:polar-latency-fer-placeholder}
  \end{minipage}\hfill
  \begin{minipage}[t]{0.48\linewidth}
    \centering
    \vspace{0pt}
    \begin{minipage}[c][0.94\linewidth][c]{\linewidth}
      \centering
      \resizebox{0.98\linewidth}{!}{\begin{tikzpicture}[
  x=0.72cm,y=0.72cm,
  font=\scriptsize,
  >={Latex[length=1.6mm]},
  rv/.style={circle,draw=black!60,fill=white,minimum size=5.7mm,inner sep=0pt},
  fwd/.style={circle,draw=orange!85!black,fill=orange!10,minimum size=3.5mm,inner sep=0pt},
  rev/.style={circle,draw=teal!70!black,fill=teal!10,minimum size=3.5mm,inner sep=0pt},
  physical/.style={draw=blue!65!black,fill=blue!5,rounded corners=3pt,align=center,inner sep=3pt},
  fa/.style={-{Latex},draw=orange!85!black,thick},
  ra/.style={-{Latex},draw=teal!70!black,thick},
  pa/.style={-{Latex},draw=blue!65!black,thick},
  ca/.style={-{Latex},draw=blue!65!black,dashed}
]
\node[rv] (x) at (0,1.6) {$X$};
\node[physical] (ch) at (3.1,1.6) {$P_{R\mid X}$\\physical channel};
\node[rv,draw=blue!65!black] (r) at (6.2,1.6) {$R$};
\draw[pa] (x)--(ch);
\draw[pa] (ch)--(r);

\node[fwd] (f0) at (0,3.3) {};
\node[fwd] (f1) at (2.0,3.3) {};
\node[fwd] (f2) at (4.2,3.3) {};
\node[fwd] (f3) at (6.2,3.3) {};
\draw[fa] (x)--(f0);
\draw[fa] (f0)--(f1);
\draw[fa] (f1)--node[above] {$\cdots$}(f2);
\draw[fa] (f2)--(f3);
\node[below=2.2mm,text=orange!85!black] at (f1.south) {forward masking};
\node[above=1mm] at (f0) {$Y(t^{(0)})$};
\node[above=1mm] at (f1) {$Y(t^{(1)})$};
\node[above=1mm] at (f2) {$Y(t^{(K-1)})$};
\node[above=1mm] at (f3) {$Y(t^{(K)})$};

\node[rev] (q0) at (0,-1.2) {};
\node[rev] (q1) at (2.0,-1.2) {};
\node[rev] (q2) at (4.2,-1.2) {};
\node[rev] (q3) at (6.2,-1.2) {};
\draw[ra] (q3)--(q2);
\draw[ra] (q2)--node[below] {$\cdots$}(q1);
\draw[ra] (q1)--(q0);
\node[below=5.8mm,text=teal!70!black] at (q1.south) {decoder};
\node[above left=0.8mm and -0.5mm] at (q0) {$\widetilde Y^{(0)}$};
\node[above left=0.8mm and -0.5mm] at (q1) {$\widetilde Y^{(1)}$};
\node[above left=0.8mm and -0.5mm] at (q2) {$\widetilde Y^{(K-1)}$};
\node[above left=0.8mm and -0.5mm] at (q3) {$\widetilde Y^{(K)}$};
\node[below=1.5mm,align=center] at (q0) {output\\$\widehat X$};
\node[below=1.5mm,align=center] at (q3) {base\\sample};
\coordinate (rbusright) at (6.85,0.35);
\coordinate (rbusleft) at (0,0.35);
\draw[draw=blue!65!black,dashed,thick] (r.east)
  .. controls (7.25,1.6) and (7.25,0.35) .. (rbusright);
\draw[draw=blue!65!black,dashed,thick] (rbusright)--(rbusleft);
\node[above=1.2mm,text=blue!65!black] at (3.4,0.35)
  {fixed $R$ throughout};
\draw[ca] (0,0.35)--(q0.north);
\draw[ca] (2.0,0.35)--(q1.north);
\draw[ca] (4.2,0.35)--(q2.north);
\draw[ca] (6.2,0.35)--(q3.north);
\end{tikzpicture}}
      \vspace{-4mm}
    \end{minipage}
\captionof{figure}{Conditional masking-diffusion decoding.
The observation $R$ remains fixed while an auxiliary process masks $X$; the decoder reverses masking to estimate $\widehat X$.}
    \label{fig:conditional-diffusion-process}
  \end{minipage}
\end{figure}

Diffusion-based decoders provide a natural starting point, having shown competitive error-correction performance through iterative denoising and score-based inference \citep{choukrounDenoisingDiffusionError2023,helvitsScoreBasedError2026}.
We first model the decoder as a conditional diffusion process based on conservation laws developed for coding \citep{meassonConservationLawsCoding2006} and extended to diffusion under model mismatch \citep{aharoniConservationLawsDiffusion2026}. 
In our formulation, the physical-channel observation remains fixed as side information while a separate auxiliary process corrupts the codeword with a broad class of memoryless auxiliary noise processes. The decoder then follows the reverse of this auxiliary process to recover the codeword having the physical observation as a fixed condition, as illustrated in Figure~\ref{fig:conditional-diffusion-process}.
Formally, this process should generate a sample from the posterior distribution of transmitted codeword given the physical observation.
\revision2{In typical communication scenarios, the posterior is often sharply concentrated on a single codeword and thus approximates a point mass. In this regime, sampling from the posterior is approximately equivalent to selecting its argmax.}
For the purpose of developing a \revision2{beam} decoder, we choose masking noise as the auxiliary process, even though the conservation laws in \citep{meassonConservationLawsCoding2006,aharoniConservationLawsDiffusion2026} accommodate a broad class of memoryless auxiliary noise processes.

\par \henrytask{8}{Having established the conditional diffusion framework, we then augment the masking-diffusion decoder with guessing, inspired by the Maxwell construction of \citet{meassonMaxwellConstructionHidden2008}.}
Specifically, at every reverse decision, the decoder branches on multiple bit assignments and scores each path using the log probabilities of the reverse transitions along that path.
When the number of paths exceeds the \revision2{beam} size, the decoder prunes the least likely paths based on the accumulated reverse-transition log probabilities. \revision2{Upon termination, the decoder selects the most likely codeword among the completed candidates in the beam.} 
\par

\revision2{Our experiments across several code families on BPSK-AWGN show that MBS improves on the corresponding single-trajectory conditional diffusion decoder and achieves the best decoding performance among the compared neural decoders in all evaluated scenarios. Figure~\ref{fig:polar-latency-fer-placeholder} compares our methods with classical and neural decoders for Polar$(64,32)$ at 4 dB.}

\section{Related Work}

\paragraph{Neural decoders.}
\revision2{Learned belief-propagation (BP) decoders train message weights or update rules on Tanner graphs~\citep{nachmaniLearningDecodeLinear2016,nachmaniAutoregressiveBeliefPropagation2021}. Transformer decoders such as ECCT and CrossMPT retain the parity-check structure while replacing message updates with learned attention~\citep{choukrounErrorCorrectionCode2022,parkCrossMPTCrossAttentionMessagePassing2025}. Neural polar decoders also combine learned inference with list decoding, although sequential bit decisions limit parallelism~\citep{aharoniDatadrivenNeuralPolar2024,hirschStudyNeuralPolar2025}. Diffusion-based decoders, including DDECC and SB-ECC, recover codewords through iterative denoising of Gaussian channel noise~\citep{choukrounDenoisingDiffusionError2023,helvitsScoreBasedError2026}. Our conditional formulation instead holds the physical observation fixed and introduces a separate auxiliary masking process. This separation allows the learned posterior estimator to condition on partial bit assignments and support beam search over decoding trajectories.}

\paragraph{Decimation and beam search.}
\revision2{Guessing and retaining alternative decoding paths have a long history, including the Maxwell decoder~\citep{meassonMaxwellConstructionHidden2008} and successive-cancellation list decoding~\citep{talListDecodingPolar2015}. More directly related, \citet{buchbergerLearnedDecimationNeural2021} combine least-reliable-bit branching with learned decimation for neural BP on classical LDPC codes. For quantum LDPC codes, \citet{yeBeamSearchDecoder2025} use BP-guided branching and reliability-based beam pruning, while \citet{aydinSoftDecodingQuantum2026} use cumulative BP iterations for post-selection. Our method shares the use of branching to improve an approximate decoder. We approach this mechanism through conditional masked diffusion: posterior marginals guide bit assignments, accumulated reverse-transition log probabilities rank paths, and parity constraints eliminate infeasible candidates. The connection to Maxwell decoding therefore complements the more recent decimation and beam-search literature. Extended related work appears in Appendix~\ref{app:extended-related-work}.}

\section{Preliminaries and Background}
\label{sec:preliminaries}

\subsection{Error-correcting codes}
\label{sec:prelim-code-channel}

We consider the standard coded transmission setup using a binary linear block code $\mathcal C$.
The code is specified by a generator matrix $G\in\{0,1\}^{n\times k}$ and a parity-check matrix $H\in\{0,1\}^{r\times n}$ such that $HG=\mathbf{0}_{r\times k}$ over $\mathbb F_2$ \citep{richardsonModernCodingTheory2008}.
The symbol $\mathbf{0}_{r\times k}$ denotes an all-zero matrix of the indicated dimensions.
For an $(n,k)$ code, $\operatorname{rank}(H)=n-k$ and the number $r$ of listed parity checks satisfies $r\geq n-k$, with equality when the checks are linearly independent.
Given a message $m\in\{0,1\}^k$, the encoder produces the length-$n$ codeword
\begin{equation}
  x=Gm\in\mathcal C\subseteq\{0,1\}^n,
  \qquad Hx=0,
  \label{eq:linear-code-definition}
\end{equation}
and the code rate is $k/n$.
For transmission, we map the codeword to its binary phase-shift-keying (BPSK) representation $x_{\mathrm s}=\mathbf 1-2x\in\{\pm1\}^n$.
We use uppercase $X$ for the random transmitted codeword and lowercase $x$ for one realization.
\par

\subsection{Physical observation and auxiliary noise}
\label{sec:prelim-auxiliary-path}

We distinguish two channels because they play different roles in the decoder.
The physical channel produces the observation $R$, which remains fixed and conditions every decoding decision.
The auxiliary channel instead governs the evolving state $Y(t)$ and serves as an internal mechanism for taking decisions about the transmitted codeword.

\paragraph{Physical observation.}
The physical channel is described by the conditional distribution \mbox{$P_{R\mid X}(r\mid x)$}, from a transmitted codeword $x\in\mathcal C$ to a received observation $R$.
This distribution is set by the communication medium.
The realization $R=r$ is the physical information available to the decoder and remains fixed throughout decoding.

\paragraph{Auxiliary noise.}
Separately, the decoding process is designed by choosing an auxiliary scalar noising family $\mathcal W=\{W_t\}_{t\in\mathcal T}$, where $\mathcal T\subseteq\mathbb R_{\geq 0}$ is the scalar noise-parameter domain.
This family facilitates the decision process of the decoder.
We use only the homogeneous auxiliary channel: at noise level $t$, every coordinate is passed through the same scalar channel.
Thus $Y(t)=(Y_1(t),\ldots,Y_n(t))$ has conditional distribution
\begin{equation}
  W_t^{\otimes n}(y\mid x)
  =\prod_{i=1}^n W_t(y_i\mid x_i),
  \qquad t\in\mathcal T.
  \label{eq:auxiliary-product-channel}
\end{equation}
\henrytask{3}{\revision2{We assume that $\mathcal W$ is degraded: for every $t'>t$, there is a channel $Q_{t,t'}(y\mid\tilde y)$ on the auxiliary output alphabet $\mathcal Y$ satisfying
\[
  W_{t'}(y\mid x)
  =\sum_{\tilde y\in\mathcal Y}W_t(\tilde y\mid x)\,Q_{t,t'}(y\mid\tilde y).
\]
}
Assume that $\mathcal T$ has a clean endpoint $t^{\mathrm{cln}}$ and an uninformative endpoint $t^{\mathrm{ind}}$, for which $Y(t^{\mathrm{cln}})=X$ almost surely and $Y(t^{\mathrm{ind}})\perp X$. \revision2{For example, for masking noise with mask probability $t\in[0,1]$, the clean endpoint is $t^{\mathrm{cln}}=0$, where $Y(0)=X$, and the uninformative endpoint is $t^{\mathrm{ind}}=1$, where $Y(1)=\ast^n$ is the all-erasure word.}
The scalar parameter $t$ therefore indexes the decoder trajectory from clean to uninformative.
\revision2{For general auxiliary channels, regularity conditions are given by \citet{meassonGeneralizedAreaTheorem2009,aharoniConservationLawsDiffusion2026}. Appendix~\ref{app:conditional-conservation-laws} proves the masking identities used here directly.}
\revision2{Because the auxiliary channel family is degraded and its randomness is independent of the physical observation conditional on $X$, coupling $Y(t')$ to $Y(t)$ through the degradation channel yields the Markov chain}
\begin{equation}
  \revision2{R\longleftrightarrow X\longleftrightarrow Y(t)\longleftrightarrow Y(t'),\qquad t'>t.}
  \label{eq:conditional-diffusion-markov}
\end{equation}}
\par

\section{Decoding Error-Correcting Codes as Conditional Diffusion}
\label{sec:conditional-diffusion-decoder}
This section commences with a brief background of the conservation laws for masked diffusion, which connect the local bitwise posterior quantities to the global conditional entropy of the codeword given the physical observation. Then, it describes how these posterior marginals define the diffusion decoder and its \revision2{MBS} extension.
\par
\subsection{Conservation law for masked diffusion}
\label{sec:masked-conservation-law}

\par \henrytask{5}{\revision2{Consider a random vector $X$ and its masked version $Y(t)$, formed by independently replacing each bit with the erasure symbol $\ast$ with probability $t\in[0,1]$. Let $Y_{-i}(t)$ denote all coordinates of $Y(t)$ except the $i$th. The area theorem of \citet{meassonGeneralizedAreaTheorem2009} expresses the entropy of $X$ through derivatives along this masking path rather than through the chain rule over variables.}}

\par \henrytask{5}{\revision2{Applying the area theorem to the masking channel with the physical-channel observation $R$ held fixed as side information, the derivative of $H(X\mid \revision2{Y(t),R})$ decomposes into bitwise \mbox{conditional entropies:}}}
\begin{equation}
  \frac{d}{dt}H(X\mid \revision2{Y(t),R})
  =\sum_{i=1}^n
    H(X_i\mid \revision2{Y_{-i}(t),R}).
  \label{eq:conditional-masked-gexit}
\end{equation}
\revision2{By the fundamental theorem of calculus, integrating from $t=0$ to $t=1$ connects the local bitwise conditional entropies to the global conditional entropy $H(X\mid R)$:}
\begin{equation}
  H(X\mid \revision2{Y(1),R}) - H(X\mid \revision2{Y(0),R}) = \int_0^1\sum_{i=1}^n
    H(X_i\mid \revision2{Y_{-i}(t),R})\,dt,
  \label{eq:conditional-masked-conservation}
\end{equation}
which equals $H(X\mid R)$.

\par \henrytask{6}{\revision2{We establish a conditional conservation law for mismatched cross-entropy, extending the unconditional result of \citet{aharoniConservationLawsDiffusion2026} to include the physical-channel observation $R$ as side information.}} \revision2{Let $P_X$ and $Q_X$ be the true and model distributions on codewords. The joint laws $P_{X,R}=P_XP_{R\mid X}$ and $Q_{X,R}=Q_XP_{R\mid X}$ use the same physical channel and induce the posteriors $P_{X\mid R}$ and $Q_{X\mid R}$, respectively.}
\revision2{Pass both laws of $X$ through the same masking channel $W_t$, with the auxiliary randomness conditionally independent of $R$ given $X$, so that both satisfy the Markov relation in~\eqref{eq:conditional-diffusion-markov}. Denote by} $Q_{X_i\mid \revision2{Y_{-i}(t),R}}$ the resulting model posterior of bit $i$ given $R$ and the unmasked coordinates.
Define
\begin{equation}
  \revision2{\mathcal C_{P,Q}(t):=\sum_{i=1}^n\operatorname{CE}\!\left(P_{X_i\mid Y_{-i}(t),R},Q_{X_i\mid Y_{-i}(t),R}\right),}
  \label{eq:conditional-masked-ce-integrand}
\end{equation}
where $\operatorname{CE}(P_{X\mid R},Q_{X\mid R}):=-\mathbb E_{(X,R)\sim P_{X,R}}[\log Q_{X\mid R}(X\mid R)]$ is the conditional cross-entropy. \revision2{Our conditional conservation law states that}
\begin{equation}
  \operatorname{CE}\!\left(P_{X\mid R},Q_{X\mid R}\right)
  =\int_0^1\mathcal C_{P,Q}(t)\,dt.
  \label{eq:conditional-mismatched-masked-conservation}
\end{equation}
\par

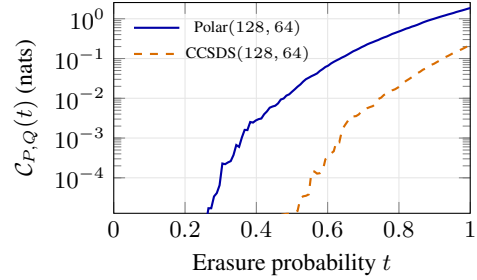
\begin{wrapfigure}{r}{0.46\textwidth}
\vspace{-\intextsep}
\footnotesize
\centering
\begin{tikzpicture}
\begin{axis}[
  width=0.98\linewidth,height=0.68\linewidth,
  xmin=0,xmax=1,xtick={0,0.2,0.4,0.6,0.8,1},
  ymode=log,ymin=1.28e-5,ymax=2.56,ytick={1e-4,1e-3,1e-2,1e-1,1},
  xlabel={Erasure probability $t$},
  ylabel={$\mathcal C_{P,Q}(t)$ (nats)},
  tick label style={font=\footnotesize},
  label style={font=\small},
  grid=major,grid style={gray!20},
  legend style={font=\tiny,draw=none,fill=none,at={(0.03,0.97)},anchor=north west},
  no markers,
]
\addplot[blue!65!black,thick]
  table[x=erasure_rate,y expr={128*\thisrow{exit_nats}},col sep=comma]{figures/data/polar12864-exit-snr4.csv};
\addlegendentry{Polar$(128,64)$}
\addplot[orange!85!black,thick,dashed]
  table[x=erasure_rate,y expr={128*\thisrow{exit_nats}},col sep=comma]{figures/data/ccsds12864-exit-snr4.csv};
\addlegendentry{CCSDS$(128,64)$}
\end{axis}
\end{tikzpicture}
\caption{\henrytask{2}{Conditional cross-entropy $\mathcal C_{P,Q}(t)$ under masked diffusion for Polar$(128,64)$ and CCSDS$(128,64)$ with AWGN side information at $E_b/N_0=4$ dB.}}
\label{fig:polar-masking-motivation}
\end{wrapfigure}

\revision2{We prove this conditional identity in Appendix~\ref{app:conditional-conservation-laws}.}
\par

\noindent\textbf{A motivating example.} Figure~\ref{fig:polar-masking-motivation} compares Polar$(128,64)$~\citep{arikanChannelPolarization2009} and CCSDS$(128,64)$~\citep{ccsdsShortBlockLDPC2015} with side information $R$ from an AWGN channel at $E_b/N_0=4$ dB. Integrating the learned curves and converting to bits gives cross-entropy estimates for the full $128$-bit codeword of $0.38$ bits for Polar and $0.028$ bits for CCSDS. The curves also suggest that Polar is harder to decode: as $t$ decreases, its prediction loss becomes negligible near $t\approx0.25$, compared with $t\approx0.5$ for CCSDS. Thus, Polar requires more coordinates to be revealed before the remaining bits become predictable. This motivates our \revision2{MBS} in Section~\ref{sec:maxwell}.\par

\par
\WFclear

\subsection{Learning posterior marginals}
\label{sec:fixed-side-information}
\label{sec:decoder-architecture}
We model the posterior marginals $Q_{X_i\mid \revision2{Y(t),R}}$ using a neural network. The conservation law requires that $P$ and $Q$ be compatible with the same auxiliary channel $W_t$: both posterior families must be induced by passing their respective codeword distributions through $W_t^{\otimes n}$. In particular, their marginals must obey the following Bayes factorization \revision2{for all $i$ and $t$}:
\begin{equation}
Q_{X_i\mid \revision2{Y(t),R}}(x_i \mid \revision2{y,r}) = \frac{Q_{X_i\mid \revision2{Y_{-i}(t),R}}(x_i \mid \revision2{y_{-i},r}) W_t(y_i|x_i)}{\sum_{x^\prime} Q_{X_i\mid \revision2{Y_{-i}(t),R}}(x^\prime \mid \revision2{y_{-i},r})W_t(y_i|x^\prime)},
\end{equation}
For the masking channel $W_t$, this becomes:
\begin{equation}
  Q_{X_i\mid \revision2{Y(t),R}}(x_i\mid \revision2{y,r})
  =\begin{cases}
    Q_{X_i\mid \revision2{Y_{-i}(t),R}}(x_i\mid \revision2{y_{-i},r})
      & y_i=\ast,\\
    \mathbf 1\{x_i=y_i\}, & y_i\in\{0,1\}
  \end{cases}.
  \label{eq:masked-bitwise-posterior}
\end{equation}
Thus, the model predicts posteriors at masked positions and is deterministic at revealed positions.
\par

\par Each training step begins by sampling a codeword and one physical channel observation, $(X,R)\sim P_XP_{R\mid X}$.
After observing $R=r$, we hold this realization fixed.
We then sample a mask probability $t\sim\operatorname{Unif}(0,1)$ and generate only the auxiliary state $Y(t)\sim W_t^{\otimes n}(\,\cdot\mid X)$.
We train the network to minimize the average cross-entropy over masked bits
\begin{equation}
  \mathcal L
  =\mathbb E_{\substack{t\sim \operatorname{Unif}(0,1), X\sim P_X \\
       R\sim P_{R\mid X}, Y(t)\sim W_t^{\otimes n}}}
    \!\left[-\frac{1}{|\mathcal M|}\sum_{i\in \mathcal M}
    \log Q_{X_i\mid \revision2{Y_{-i}(t),R}}\!\left(X_i\mid
      \revision2{Y_{-i}(t),R}\right)\right],
  \label{eq:posterior-marginal-training-loss}
\end{equation}
where $\mathcal M = \left\{i \::\:Y_i(t)=\ast\right\}$ with the requirement that $\left|\mathcal M\right| \ge 1$\footnote{During training, if none of the symbols are masked, then one random symbol in the codeword is masked.}.
\par
\subsection{Inference}
\label{sec:conditional-reverse-decoding}

\ifdefined\arxivlayout\else
\ifdefined\arxivlayout
  \begin{wrapfigure}[12]{R}{0.51\textwidth}
  \vspace{\dimexpr-\intextsep+6.7pt\relax}
  \centering
  \begin{minipage}{\linewidth}
\else
\begin{wrapfigure}[10]{r}{0.51\textwidth}
  \vspace{\dimexpr-\intextsep-\baselineskip\relax}
  \centering
  \begin{minipage}{\linewidth}
\fi
    \footnotesize
    \hrule\vspace{0.25em}
    \refstepcounter{algorithm}\label{alg:conditional-reverse-sampling}
    \textbf{Algorithm \thealgorithm}\quad Confidence-scheduled decoding
    \vspace{0.2em}\hrule\vspace{0.2em}
    \begin{algorithmic}[1]
      \Require $r$, parity-check matrix $H$,
      \Statex \hspace{\algorithmicindent}counts $(b_K,\ldots,b_1)$ with $\sum_k b_k=n$
      \State $\widetilde y^{(K)}\gets\ast^n$
      \For{$k=K,K-1,\ldots,1$}
        \State $t^{(k)}\gets$ current masked fraction
        \State Compute $\lambda_i^{(k)}$ with \eqref{eq:decoder-predicted-llr}
        \State $\widehat x_i\gets\mathbf1\{\lambda_i^{(k)}\geq0\}$, $i=1,\ldots,n$
        \State \textbf{if} $H\widehat x=0$ over $\mathbb F_2$ \textbf{then return} $\widehat x$
        \State $\mathcal G_k\gets b_k$ \revision2{masked indices with largest $|\lambda_i^{(k)}|$}
        \State $\widetilde y_i^{(k-1)}\gets\mathbf 1\{\lambda_i^{(k)}\geq0\}$ for $i\in\mathcal G_k$
      \EndFor
      \State \Return $\widehat x=\widetilde y^{(0)}$
    \end{algorithmic}
    \vspace{0.2em}\hrule
  \end{minipage}
\end{wrapfigure}

\fi

\par During inference, the learned bitwise posterior defines the reverse transitions.
We describe conditional masked diffusion in Section~\ref{sec:cond-diffusion} and its \revision2{MBS} extension in Section~\ref{sec:maxwell}.
\par

\subsubsection{Conditional-diffusion decoding.}\label{sec:cond-diffusion}
Before decoding, we specify the number of bits $b_K,\ldots,b_1$ to reveal at each of $K$ steps, with $\sum_{k=1}^K b_k=n$.
\revision2{We call the schedule \emph{confidence-balanced} when the reveal counts are equal up to rounding and each step reveals the masked bits with largest absolute LLRs.}
\ifdefined\arxivlayout
\par
\fi
The decoder starts from the all-erasure word $\widetilde y^{(K)}=\ast^n$ and keeps the physical observation $R=r$ fixed.
At step $k$, the current word is $\widetilde y^{(k)}$ and the noise level $t^{(k)}=(b_k+\cdots+b_1)/n$ is the fraction of bits still masked.
One forward pass evaluates the posterior marginals in \eqref{eq:masked-bitwise-posterior} for this state.
For each bit $i$, we compute the predicted log-likelihood ratio (LLR)
\begin{equation}
  \lambda_i^{(k)}
  =\log\frac{Q_{X_i\mid \revision2{Y(t^{(k)}),R}}(1\mid \revision2{\widetilde y^{(k)},r})}
                 {Q_{X_i\mid \revision2{Y(t^{(k)}),R}}(0\mid \revision2{\widetilde y^{(k)},r})},
  \label{eq:decoder-predicted-llr}
\end{equation}
and $|\lambda_i^{(k)}|$ measures confidence.
\par
After each forward pass, we form the early hard estimate $\widehat x_i=\mathbf1\{\lambda_i^{(k)}\geq0\}$ for every $i$. If $H\widehat x=0$ over $\mathbb F_2$, we return $\widehat x$ immediately. Otherwise, we continue with the scheduled reveal step.
We select the $b_k$ still-masked positions with the largest $|\lambda_i^{(k)}|$ to form $\mathcal G_k$, then reveal them using the LLR hard decision induced by the marginal backward kernel
\par
\ifnum\value{WF@wrappedlines}>1
  \vspace{\dimexpr\value{WF@wrappedlines}\baselineskip-\baselineskip\relax}
\fi
\WFclear
\begin{equation}
  Q_{\widetilde Y_i^{(k-1)}\mid \revision2{\widetilde Y^{(k)},R}}(\widetilde y_i^{(k-1)}\mid \revision2{\widetilde y^{(k)},r})
  =
  \begin{cases}
    Q_{X_i\mid \revision2{Y(t^{(k)}),R}}(\widetilde y_i^{(k-1)}\mid \revision2{\widetilde y^{(k)},r}),
      & i\in\mathcal G_k,\\
    \mathbf 1\{\widetilde y_i^{(k-1)}=\widetilde y_i^{(k)}\}, & i\notin\mathcal G_k.
  \end{cases}
  \label{eq:decoder-masking-reverse-step}
\end{equation}
For $i\in\mathcal G_k$, the new value is $\widetilde y_i^{(k-1)}=\mathbf 1\{\lambda_i^{(k)}\geq0\}$; every other coordinate keeps its current value.
If no early exit occurs, after $K$ steps all bits are revealed and the noise level has decreased from $t^{(K)}=1$ to $t^{(0)}=0$.
Algorithm~\ref{alg:conditional-reverse-sampling} summarizes the procedure. The next section describes the \revision2{MBS}, which augments Algorithm~\ref{alg:conditional-reverse-sampling} with a \revision2{beam} of candidate paths.
\par
\ifnum\value{WF@wrappedlines}>1
  \vspace{\dimexpr\value{WF@wrappedlines}\baselineskip-\baselineskip\relax}
\fi
\WFclear

\subsubsection{\texorpdfstring{\revision2{Maxwell beam search}}{Maxwell beam search}}\label{sec:maxwell}
The Maxwell decoder was introduced for erasure decoding to bridge belief propagation (BP) and maximum a posteriori (MAP) decoding \citep{meassonMaxwellConstructionHidden2008}.
When BP cannot resolve an erased bit, it branches on both possible values and uses parity constraints to eliminate inconsistent hypotheses.
Our formulation likewise recovers erased bits, with the physical observation $R$ providing additional side information.
We adapt this branching mechanism by using the learned neural posterior marginals in place of BP predictions and retaining a finite beam of candidate decoding paths.

\paragraph{Initialization.}
\ifdefined\preprintupdatedresults
The decoder takes the observation $r$, parity-check matrix $H$, beam width $L$, a maximum of $\revision2{K_B}$ branching rounds, an initial branching depth $d_0$, and a completion budget of $\revision2{K_C}$ steps per attempt. We require $1\leq d_0\leq \revision2{K_B}$ and $2^{d_0}\leq L$.
\else
The decoder takes the observation $r$, parity-check matrix $H$, beam width $L$, a maximum of $\revision2{K_B}=16$ branching rounds, an initial branching depth $d_0=\log_2 L$, and a completion budget of $\revision2{K_C}=8$ steps per attempt.
We take $L$ to be a power of two with $1\leq\log_2 L\leq \revision2{K_B}$.
\fi
\revision2{The beam contains a list of candidate--score tuples and is initialized as $\mathcal B_{K_B}=\{(\ast^n,0)\}$, consisting of one all-erasure word with path score zero.}
At step $k$, the beam contains $L_k=|\mathcal B_k|\leq L$ candidates $(\widetilde y_\ell^{(k)},s_\ell^{(k)})$, indexed by $\ell=1,\ldots,L_k$.
Evaluating the beam $\mathcal B_k$ gives the LLRs $\{\lambda_{\ell,i}^{(k)}\}$, for $\ell=1,\ldots,L_k$ and $i=1,\ldots,n$, where $\lambda_{\ell,i}^{(k)}$ is defined by \eqref{eq:decoder-predicted-llr}. Initially, $L_{K_B}=1$ and $t_1^{(K_B)}=1$.

\paragraph{\texorpdfstring{\revision2{Greedy exit.}}{Greedy exit.}}
The first completion attempt uses the prediction from the all-erasure state and $r$. Using these predicted LLRs, it forms the greedy word
\begin{equation}
  x_i^{\mathrm{gr}}=\mathbf1\{\lambda_{1,i}^{(K_B)}\geq0\},\qquad i=1,\ldots,n.
  \label{eq:maxwell-final-completion}
\end{equation}
If $Hx^{\mathrm{gr}}=0$, it returns this word immediately. Otherwise, it invokes the branching \mbox{mechanism below.}

\paragraph{\texorpdfstring{\revision2{Initial branching.}}{Initial branching.}}
\ifdefined\preprintupdatedresults
If this initial attempt fails, the decoder performs $d_0$ sequential binary branching rounds. Each round recomputes the predictions.
\else
If this initial attempt fails, the decoder performs $d_0=\log_2 L$ sequential binary branching rounds. Each round recomputes the predictions.
\fi For each candidate $\ell$, let $i_{\ell,k}=\operatorname*{arg\,min}_{i:\widetilde y_{\ell,i}^{(k)}=\ast}|\lambda_{\ell,i}^{(k)}|$ be its least-confident masked position and $\bar y_{\ell,\,i_{\ell,k}\leftarrow b}^{(k-1)}$ the child obtained \revision2{from $\widetilde y_\ell^{(k)}$} by assigning $b\in\{0,1\}$ at that position. Parity propagation $\operatorname{Prop}_H$ then fixes any masked position participating in a parity check whose other positions are all revealed; if a contradiction is detected, the child is discarded. \revision2{For each surviving child, define its propagated state once as $\widetilde y_{\ell,b}^{(k-1)}:=\operatorname{Prop}_H(\bar y_{\ell,\,i_{\ell,k}\leftarrow b}^{(k-1)})$.} Its score is computed using only the assigned bit $b$, with no contribution from parity-forced bits\footnote{\revision2{Parity-forced bits are determined uniquely by the current assignments and parity constraints, so their conditional probability is one and their log-probability contribution is $\log 1=0$.}}:
\begin{equation}
  s_{\ell,b}^{(k-1)}
  =s_\ell^{(k)}+
  \log Q_{\widetilde Y_{i_{\ell,k}}^{(k-1)}\mid \revision2{\widetilde Y^{(k)},R}}
  \!\left(b\mid \revision2{\widetilde y_\ell^{(k)},r}\right),\qquad b\in\{0,1\}.
  \label{eq:maxwell-beam-score}
\end{equation}
\revision2{The surviving propagated children $\widetilde y_{\ell,b}^{(k-1)}$, together with their scores $s_{\ell,b}^{(k-1)}$, form the next beam $\mathcal B_{k-1}$.}
Fully revealed parents are carried over unchanged.

\begin{figure}[!t]
  \centering
  \definecolor{mxink}{HTML}{233750}
  \definecolor{mxblue}{HTML}{426FE0}
  \definecolor{mxamber}{HTML}{C98520}
  \definecolor{mxteal}{HTML}{168777}
  \definecolor{mxpurple}{HTML}{7857A6}
  \definecolor{mxgray}{HTML}{A5AFBD}
  \newcommand{\mxword}[3]{%
    \foreach \bit/\kind [count=\j from 0] in {#3}{%
      \pgfmathsetlengthmacro{\mxshift}{\j*5.2pt}%
      \node[bit,\kind,anchor=west,outer sep=0pt]
        at ([xshift=\mxshift]#1,#2) {\bit};}}
  \begin{tikzpicture}[
    x=\dimexpr\textwidth/400\relax,y=1.3pt,
    font=\sffamily\fontsize{7.5}{9}\selectfont,text=mxink,
    bit/.style={minimum width=5.2pt,minimum height=8pt,inner sep=0pt,
      font=\sffamily\bfseries\fontsize{7.5}{8}\selectfont},
    m/.style={draw=mxgray!60,fill=mxgray!7,text=mxgray,rounded corners=1pt},
    v/.style={draw=mxink!45,fill=mxink!12,rounded corners=1pt},
    b/.style={draw=mxamber!65,fill=mxamber!16,rounded corners=1pt},
    c/.style={draw=mxamber,fill=mxamber!55,rounded corners=1pt},
    f/.style={draw=mxteal!70,fill=mxteal!20,rounded corners=1pt},
    d/.style={draw=mxblue!65,fill=mxblue!15,rounded corners=1pt},
    flow/.style={-{Latex[length=1.5mm]},draw=mxgray,line width=0.6pt},
    good/.style={flow,draw=mxteal!85!black},
    score/.style={font=\sffamily\fontsize{7}{8}\selectfont,anchor=east},
    head/.style={font=\sffamily\bfseries\fontsize{9}{10}\selectfont}]
    \path[use as bounding box] (0,-3) rectangle (380,140);
    \node[head,text=mxblue] at (75,134) {Predict};
    \node[head,text=mxteal,font=\sffamily\bfseries\fontsize{8.5}{10}\selectfont]
      at (191,134) {Diffusion completion};
    \node[head,text=mxteal] at (272,134) {Parity check};
    \node[head,text=mxteal] at (340,134) {Decision};

    \draw[rounded corners=3pt,draw=mxblue!45,fill=mxblue!2]
      (0,59) rectangle (53,115);
    \node[text=mxblue,font=\sffamily\bfseries\fontsize{7}{8}\selectfont]
      at (26.5,124) {Input beam $\mathcal B_k$};
    \node at (5,105) {A};
    \node at (5,75) {B};
    \mxword{17}{105}{*/m,*/m,*/m,1/v,*/m,*/m}
    \mxword{17}{75}{*/m,*/m,*/m,0/v,*/m,*/m}
    \node[font=\sffamily\fontsize{7}{8}\selectfont] at (32,95) {$s=-0.69$};
    \node[font=\sffamily\fontsize{7}{8}\selectfont] at (32,65) {$s=-0.69$};
    \node[circle,draw=mxblue!65,fill=mxblue!7,minimum size=26pt,inner sep=0pt]
      (model) at (75,90) {};
    \draw[flow,draw=mxblue] (53,105) -- (65,99);
    \draw[flow,draw=mxblue] (53,75) -- (65,81);
    \foreach \a in {83,90,97}
      \foreach \c in {83,90,97}
        \draw[mxblue!30,line width=.45pt] (68,\a) -- (75,\c);
    \foreach \a in {83,90,97}
      \foreach \c in {86,94}
        \draw[mxblue!30,line width=.45pt] (75,\a) -- (82,\c);
    \foreach \x/\y in {68/83,68/90,68/97,75/83,75/90,75/97,82/86,82/94}
      \fill[mxblue] (\x,\y) circle (1.2pt);
    \node[text=mxblue] at (75,117) {$r$};
    \draw[flow,draw=mxblue] (75,112) -- (75,104);
    \node[text=mxblue,font=\sffamily\fontsize{6.5}{8}\selectfont]
      at (75,65) {one batch};
    \draw[flow,draw=mxblue] (85,99) -- (97,105);
    \draw[flow,draw=mxblue] (85,81) -- (97,75);
    \node[text=mxblue,font=\sffamily\bfseries\fontsize{7}{8}\selectfont,align=center]
      at (117,122) {$|\mathrm{LLR}|$\\output};
    \mxword{103}{105}{/c,/m,/m,/v,/m,/m}
    \mxword{103}{75}{/m,/m,/m,/v,/c,/m}

    \draw[rounded corners=3pt,draw=mxteal!70,fill=mxteal!4,line width=.7pt]
      (154,64) rectangle (184,116);
    \node[text=mxteal,font=\sffamily\fontsize{6}{7}\selectfont]
      at (169,125) {Alg.\ 1};
    \foreach \y in {105,75}{
      \draw[good] (137,\y) -- (154,\y);
      \draw[good,-{Latex[length=1mm]}]
        (154,\y) -- (158,\y)
        plot[domain=158:180,samples=45,smooth]
          (\x,{\y+2.2*sin((\x-158)/22*720)})
        -- (193,\y);
      \fill[mxteal] (154,\y) circle (.8pt);
      \fill[mxteal] (184,\y) circle (.8pt);}
    \node[text=mxteal,align=center,font=\sffamily\fontsize{5.8}{7}\selectfont]
      at (169,90) {up to \ifdefined\preprintupdatedresults$\revision2{K_C}$\else8\fi\\steps};
    \node[text=mxteal,font=\sffamily\fontsize{6}{7}\selectfont]
      at (211,125) {Completed};
    \mxword{197}{105}{0/d,1/f,1/d,1/v,0/d,1/f}
    \mxword{197}{75}{0/d,0/f,1/d,0/v,0/d,0/f}
    \draw[mxteal!85!black,line width=.6pt] (230,105) -- (238,105) -- (238,75) -- (230,75);
    \draw[good] (238,90) -- (250,90);

    \node[text=mxteal,font=\sffamily\fontsize{6.5}{8}\selectfont]
      at (272,125) {Any candidate?};
    \begin{scope}[shift={(0,-5)}]
    \draw[rounded corners=3pt,draw=mxteal,fill=mxteal!8]
      (250,81) rectangle (294,109);
    \node[text=mxteal!85!black,font=\sffamily\bfseries\fontsize{9}{10}\selectfont]
      at (272,95) {$Hx=0$?};
    \draw[good] (294,95) -- (309,95);
    \node[text=mxteal,font=\sffamily\fontsize{5.8}{7}\selectfont]
      at (301.5,102) {yes};
    \draw[rounded corners=3pt,draw=mxteal,fill=mxteal!8]
      (309,84) rectangle (333,106);
    \node[text=mxteal!85!black,font=\fontsize{12}{13}\selectfont]
      at (321,95) {$\mathcal D$};
    \draw[good] (333,95) -- (342,95);
    \mxword{345}{95}{0/d,1/f,1/d,1/v,0/d,1/f}
    \node[text=mxteal,font=\sffamily\fontsize{7}{8}\selectfont]
      at (359,111) {$\widehat x$};
    \node[text=mxteal,font=\sffamily\fontsize{6}{7}\selectfont]
      at (359,81) {Return};
    \node[text=mxteal,align=center,font=\sffamily\fontsize{5.8}{7}\selectfont]
      at (321,72) {closest valid\\candidate};
    \end{scope}

    \draw[flow,draw=mxpurple,rounded corners=2pt]
      (272,76) -- (272,25.5) -- (262,25.5);
    \node[text=mxpurple,anchor=west,font=\sffamily\fontsize{5.8}{7}\selectfont]
      at (275,70) {no};
    \node[text=mxpurple,fill=white,inner sep=1pt,
      font=\sffamily\fontsize{5.8}{7}\selectfont]
      at (174,60) {Restore saved states and estimates};
    \begin{scope}[shift={(-54,-4)}]
    \draw[rounded corners=3pt,draw=mxpurple!45,fill=mxpurple!3]
      (140,10) rectangle (316,59);
    \node[text=mxpurple,font=\sffamily\bfseries\fontsize{7}{8}\selectfont]
      at (258,53) {Backtrack + branch};
    \node[text=mxpurple,font=\sffamily\bfseries\fontsize{6.5}{8}\selectfont]
      at (165,53) {Prune: top 2};
    \node[font=\sffamily\fontsize{6}{7}\selectfont] at (313,38.5) {A};
    \node[font=\sffamily\fontsize{6}{7}\selectfont] at (313,20.5) {B};
    \mxword{279}{38.5}{*/m,*/m,*/m,1/v,*/m,*/m}
    \mxword{279}{20.5}{*/m,*/m,*/m,0/v,*/m,*/m}
    \foreach \y in {43,34}
      \draw[flow,draw=mxamber] (276,38.5) -- (267,\y);
    \foreach \y in {25,16}
      \draw[flow,draw=mxamber] (276,20.5) -- (267,\y);
    \mxword{235}{43}{0/b,*/m,*/m,1/v,*/m,*/m}
    \mxword{235}{34}{1/b,*/m,*/m,1/v,*/m,*/m}
    \mxword{235}{25}{*/m,*/m,*/m,0/v,0/b,*/m}
    \mxword{235}{16}{*/m,*/m,*/m,0/v,1/b,*/m}
    \foreach \y/\s in {43/-0.91,34/-2.30,25/-1.20,16/-1.61}
      \node[score] at (228,\y) {$\s$};
    \draw[flow,draw=mxpurple] (198,29.5) -- (183,29.5);
    \mxword{148}{32}{0/b,*/m,*/m,1/v,*/m,*/m}
    \mxword{148}{18}{*/m,*/m,*/m,0/v,0/b,*/m}
    \node[text=mxpurple,font=\sffamily\fontsize{5.8}{7}\selectfont]
      at (163,43) {New beam $\mathcal B_{k-1}$};
    \end{scope}

    \node[rounded corners=3pt,draw=mxpurple!55,fill=mxpurple!3,
      minimum width=35pt,minimum height=16pt,inner sep=2pt,outer sep=0pt,
      text=mxpurple,align=center,font=\fontsize{7.5}{9}\selectfont]
      (retrycheck) at (26.5,25.5) {$k>0$?};
    \draw[flow,draw=mxpurple] (86,25.5) -- (retrycheck.east);
    \draw[flow,draw=mxpurple] (retrycheck.north) -- (26.5,59);
    \node[text=mxpurple,anchor=west,align=left,
      font=\sffamily\fontsize{5.8}{7}\selectfont]
      at (30,45) {yes: retry\\$k\gets k-1$};
    \draw[flow,draw=mxpurple] (retrycheck.south) -- (26.5,11);
    \node[text=mxpurple,anchor=west,font=\sffamily\fontsize{5.8}{7}\selectfont]
      at (30,15.5) {no};
    \draw[rounded corners=3pt,draw=mxpurple!55,fill=mxpurple!3]
      (0,0) rectangle (53,11);
    \node[text=mxpurple,font=\sffamily\bfseries\fontsize{6.5}{8}\selectfont]
      at (26.5,5.5) {Declare error};

    \node[bit,v] at (290,51) {1};
    \node[anchor=west,font=\sffamily\fontsize{6.5}{8}\selectfont] at (296,51) {revealed};
    \node[bit,m] at (290,41) {*} ;
    \node[anchor=west,font=\sffamily\fontsize{6.5}{8}\selectfont] at (296,41) {masked};
    \node[bit,c] at (290,31) {};
    \node[anchor=west,font=\sffamily\fontsize{6.5}{8}\selectfont] at (296,31) {branch};
    \node[bit,f] at (290,21) {0};
    \node[anchor=west,font=\sffamily\fontsize{6.5}{8}\selectfont] at (296,21) {parity-forced};
    \node[bit,d] at (290,11) {1};
    \node[anchor=west,font=\sffamily\fontsize{6.5}{8}\selectfont] at (296,11) {diffusion-revealed};
  \end{tikzpicture}
  \caption{\revision2{\mbsresult{2}} completion and retry\ifdefined\arxivlayout\else{} with beam-width two\fi. All candidates undergo
  diffusion completion (Algorithm~\ref{alg:conditional-reverse-sampling}) and parity checking.
  If any passes, $\mathcal D$ returns the valid candidate closest to $r$.
  Otherwise, restore the saved partial states, branch on their least-confident
  masked bits, and retain the top two children by path score for the next attempt.}
  \label{fig:maxwell-beam-decoder}
\end{figure}
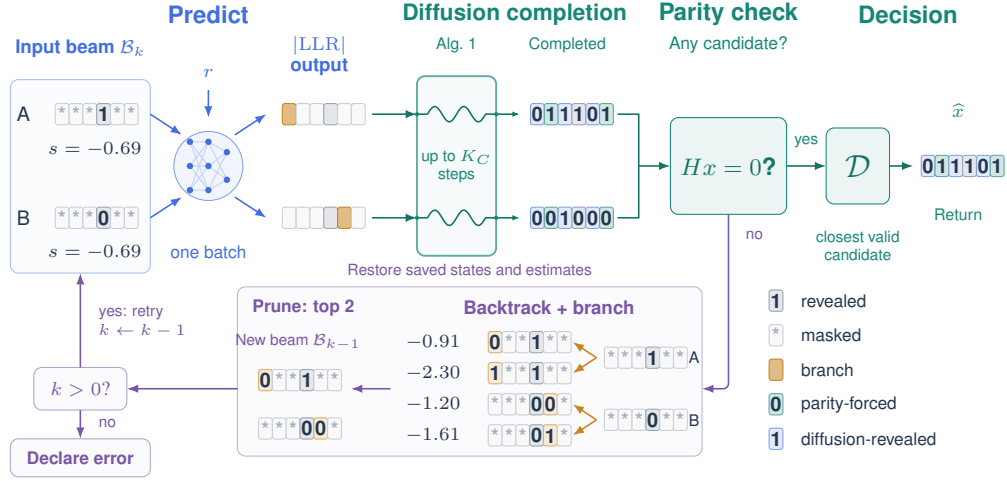

\paragraph{Main loop.}
\revision2{At each iteration, the decoder saves the current beam $\mathcal B_k$ and runs conditional diffusion on all its candidates for up to $K_C$ completion steps. At each step, it checks the hard-decision estimates using the early-exit rule of Algorithm~\ref{alg:conditional-reverse-sampling}. Let $\mathcal B^{\prime}_k$ denote the completed candidates obtained from $\mathcal B_k$ at that completion attempt. If at least one candidate in $\mathcal B^{\prime}_k$ is a valid codeword, the decoder returns one using $\mathcal D_{\mathrm{dist}}$ in~\eqref{eq:maxwell-minimum-distance-decision}.}

\revision2{During diffusion completion, we accumulate each candidate's score, initialized at $s=s_\ell^{(k)}$, over the completion steps as}
\begin{equation}
 s\leftarrow s+\sum_{i\in\mathcal G}\log\max_{b\in\{0,1\}}Q_{X_i\mid \revision2{Y(t),R}}(b\mid \revision2{\widetilde y,r}),
 \label{eq:maxwell-completion-score}
\end{equation}
\revision2{Here $\widetilde y$ is the current completion state, $t$ its masked fraction, and $\mathcal G$ the positions revealed at that step by Algorithm~\ref{alg:conditional-reverse-sampling}, including all remaining masked positions on early exit.}

\revision2{If no $x\in\mathcal B^{\prime}_k$ satisfies $Hx=0$, the decoder discards $\mathcal B^{\prime}_k$ and backtracks to $\mathcal B_k$. For each $\ell$, it branches at $i_{\ell,k}$ with $b\in\{0,1\}$, applies $\operatorname{Prop}_H$, and discards contradictions. Pruning uses the score ranking of~\eqref{eq:maxwell-score-decision}: up to $L$ surviving children with largest $s_{\ell,b}^{(k-1)}$ from~\eqref{eq:maxwell-beam-score} form $\mathcal B_{k-1}$. Completion and branching repeat for at most $K_B$ branching rounds, including the initial $d_0$. Failure is declared if the beam is empty or the final completed beam $\mathcal B^{\prime}_0$ contains no valid codeword.}
\paragraph{Codeword selection.}
We consider two metrics on $\revision2{\mathcal B^{\prime}_k}$: minimum Euclidean distance and maximum accumulated path score. The minimum-distance rule selects the most likely codeword under BPSK-AWGN, equivalently the one with minimum squared Euclidean distance:
\begin{equation}
  \mathcal D_{\mathrm{dist}}(\revision2{\mathcal B^{\prime}_k},r)
  =\operatorname*{arg\,min}_{x\in\revision2{\mathcal B^{\prime}_k}:\,\revision2{Hx=0}}
  \|r-(\mathbf1-2x)\|_2^2.
  \label{eq:maxwell-minimum-distance-decision}
\end{equation}
Alternatively, let $s(x)$ be the accumulated score of the path producing $x$. The score-based rule is
\begin{equation}
  \mathcal D_{\mathrm{score}}(\revision2{\mathcal B^{\prime}_k})
  =\operatorname*{arg\,max}_{x\in\revision2{\mathcal B^{\prime}_k}:\,\revision2{Hx=0}}s(x).
  \label{eq:maxwell-score-decision}
\end{equation}

\paragraph{Path metric.}
\revision2{With exact posterior marginals, sequential revelation yields the log-posterior score $\log P_{X\mid R}(x\mid r)$, whereas parallel revelation gives a lower bound in expectation. See Appendix~\ref{app:parallel-score-expectation} for details and the extension to learned marginals.}
\section{Experiments}
We evaluate conditional diffusion and \revision2{MBS} on standard linear block codes over BPSK-AWGN. We also examine the distribution of decoding steps and measured latency.

\subsection{Experimental setup}

\ifdefined\preprintupdatedresults
\suppressfloats[t]
\begin{table}[t]
  \centering
\caption{\textbf{BER comparison over BPSK--AWGN.} \revision2{\revision2{MBS} uses beam size $L=64$.}
Values are $-\ln(\mathrm{BER})$ (higher is better), at the indicated $E_b/N_0$ in dB.
Bold and underline mark the best and second-best distinct values after rounding to two decimals; ties share the same formatting.
The symbol $\dagger$ indicates no observed errors; the estimate assumes one frame error with $d_{\min}$ bit errors (Appendix~\ref{app:training-simulation-details}).}
  \label{tab:short-code-ber}
\footnotesize
  \setlength{\tabcolsep}{2.1pt}
  \resizebox{\textwidth}{!}{%
  \begin{tabular}{@{}ll*{6}{ccc}@{}}
    \toprule
    \multicolumn{2}{@{}l}{Architecture} &
    \multicolumn{3}{c}{Non-neural decoders} &
    \multicolumn{3}{c}{Neural decoder} &
    \multicolumn{6}{c}{Diffusion-based neural decoders} &
    \multicolumn{6}{c}{Our decoders} \\
    \cmidrule(lr){1-2}\cmidrule(lr){3-5}\cmidrule(lr){6-8}
    \cmidrule(lr){9-14}\cmidrule(lr){15-20}
    Family & Code &
    \multicolumn{3}{c}{BP} &
    \multicolumn{3}{c}{CrossMPT} &
    \multicolumn{3}{c}{DDECC} &
    \multicolumn{3}{c}{SB-ECC} & \multicolumn{3}{c}{Diffusion} & \multicolumn{3}{c}{\revision2{MBS}\revision2{, $L=64$}} \\
    \cmidrule(lr){3-5}\cmidrule(lr){6-8}\cmidrule(lr){9-11}
    \cmidrule(lr){12-14}\cmidrule(lr){15-17}\cmidrule(lr){18-20}
    & & 4 & 5 & 6 & 4 & 5 & 6 & 4 & 5 & 6 & 4 & 5 & 6 & 4 & 5 & 6 & 4 & 5 & 6 \\
    \midrule

    Polar & $(64,32)$ & 4.26 & 5.38 & 6.50 & 7.50 & 9.97 & 13.31 & 7.03 & 9.69 & 12.97 & \underline{7.77} & 10.30 & \underline{13.78} & 7.70 & \underline{10.43} & 13.57 & \revision2{\textbf{7.92}} & \revision2{\textbf{10.64}} & \revision2{\textbf{13.95}} \\
    Polar & $(64,48)$ & 4.74 & 5.94 & 7.42 & 6.51 & \underline{8.70} & \underline{11.31} & 6.07 & 8.40 & 10.90 & \underline{6.63} & 8.64 & 11.27 & 6.53 & 8.63 & 11.05 & \revision2{\textbf{6.73}} & \revision2{\textbf{8.82}} & \revision2{\textbf{11.40}} \\
    Polar & $(128,64)$ & 4.10 & 5.11 & 6.15 & 7.52 & 11.21 & 14.76 & 8.16 & 12.04 & 16.27 & \underline{9.03} & \underline{13.13} & \underline{16.94} & 7.62 & 11.04 & 14.75 & \revision2{\textbf{9.51}} & \revision2{\textbf{13.32}} & \revision2{\textbf{17.36}} \\
    Polar & $(128,86)$ & 4.49 & 5.65 & 6.97 & 7.51 & 10.83 & 15.24 & 7.92 & 11.34 & 15.61 & \underline{8.03} & \underline{11.48} & \underline{16.10} & 7.68 & 11.04 & 14.57 & \revision2{\textbf{8.92}} & \revision2{\textbf{12.54}} & \revision2{\textbf{16.97}} \\
    Polar & $(128,96)$ & 4.61 & 5.79 & 7.08 & 7.15 & 10.15 & 13.13 & 7.24 & 10.32 & 13.26 & \underline{7.64} & 10.32 & \underline{13.37} & 7.36 & \underline{10.34} & 13.20 & \revision2{\textbf{8.09}} & \revision2{\textbf{10.74}} & \revision2{\textbf{13.38}} \\
    \addlinespace

    LDPC & $(49,24)$ & 6.23 & 8.19 & 11.72 & 6.68 & 9.52 & 13.19 & 6.26 & 9.07 & 12.71 & \underline{7.74} & \underline{10.88} & \underline{14.63} & 6.97 & 10.00 & 13.98 & \revision2{\textbf{8.70}} & \revision2{\textbf{11.37}} & \revision2{\textbf{15.13}} \\
    LDPC & $(121,60)$ & 5.64 & 8.88 & 13.33 & 5.74 & 9.26 & 14.78 & 6.22 & 10.18 & 15.89 & \underline{6.33} & \underline{10.38} & \underline{16.38} & 5.96 & 9.94 & 15.92 & \revision2{\textbf{8.32}} & \revision2{\textbf{13.60}} & \revision2{\textbf{21.04}} \\
    LDPC & $(121,70)$ & 6.91 & 10.66 & 15.62 & 7.06 & 11.39 & 17.52 & 7.64 & 12.30 & 17.98 & \underline{7.75} & \underline{12.69} & \underline{19.24} & 7.42 & 12.09 & 18.49 & \revision2{\textbf{10.36}} & \revision2{\textbf{15.96}} & \revision2{\textbf{22.52}} \\
    LDPC & $(121,80)$ & 7.84 & 11.86 & 17.33 & 7.99 & 12.75 & 18.15 & 8.62 & 13.55 & 18.26 & \underline{8.84} & \underline{14.02} & \underline{20.42} & 8.34 & 12.32 & 13.29 & \revision2{\textbf{11.09}} & \revision2{\textbf{16.77}} & \revision2{\textbf{23.22}} \\
    \addlinespace

    BCH & $(63,36)$ & 4.03 & 5.42 & 7.26 & 5.03 & 6.91 & 9.37 & 5.30 & 7.32 & 10.25 & \underline{5.74} & \underline{8.12} & \underline{11.20} & 5.41 & 7.53 & 10.55 & \revision2{\textbf{7.42}} & \revision2{\textbf{10.59}} & \revision2{\textbf{14.54}} \\
    BCH & $(63,45)$ & 4.36 & 5.55 & 7.26 & 5.90 & 8.20 & 11.62 & 5.67 & 8.07 & 11.41 & \underline{6.58} & \underline{9.48} & \underline{13.17} & 6.03 & 8.63 & 12.13 & \revision2{\textbf{7.74}} & \revision2{\textbf{10.99}} & \revision2{\textbf{15.53}} \\
    BCH & $(63,51)$ & 4.50 & 5.82 & 7.42 & 5.78 & 8.08 & 11.41 & 5.37 & 7.48 & 10.51 & \underline{6.19} & \underline{8.82} & \underline{12.45} & 5.87 & 8.38 & 11.76 & \revision2{\textbf{6.63}} & \revision2{\textbf{9.63}} & \revision2{\textbf{13.13}} \\
    \addlinespace
    MacKay & $(96,48)$ & 6.84 & 9.40 & 12.57 & 7.97 & 11.77 & 15.52 & 8.58 & 12.48 & 16.04 & \underline{8.97} & \underline{12.82} & \underline{16.25} & 8.51 & 12.27 & 16.02 & \revision2{\textbf{10.91}} & \revision2{\textbf{13.77}} & \revision2{\textbf{16.28}} \\
    CCSDS & $(128,64)$ & 8.00 & 12.11 & 17.17 & 7.68 & 11.88 & 17.50 & 8.40 & 13.18 & 17.47 & \underline{8.60} & \underline{13.43} & 19.20 & 8.27 & 13.23 & \underline{20.97} & \revision2{\textbf{11.46}} & \revision2{\textbf{18.07}} & \revision2{\textbf{22.94}$^{\dagger}$} \\
    \bottomrule
  \end{tabular}%
}
\end{table}

\else
\input{tables/short-code-ber}
\fi

\begin{figure}[t]
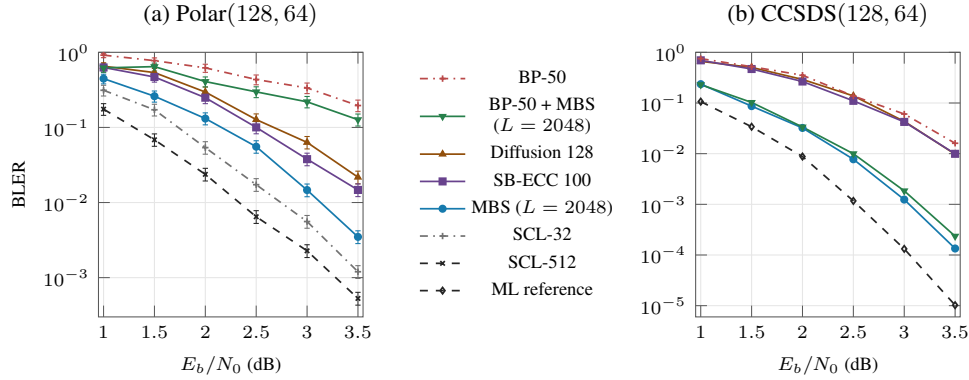

\centering
\begingroup
\definecolor{curveblue}{HTML}{167AAD}
\definecolor{curveorange}{HTML}{D78A24}
\definecolor{curvebrown}{HTML}{92500C}
\definecolor{curvelilac}{HTML}{9275B5}
\definecolor{curvepurple}{HTML}{68428D}
\definecolor{curvered}{HTML}{BA4949}
\definecolor{curvegreen}{HTML}{27804B}
\pgfplotsset{
  ours/.style={curveblue,solid,mark=*},
  plainshort/.style={curveorange,dashed,mark=diamond*},
  plainlong/.style={curvebrown,solid,mark=triangle*},
  sbshort/.style={curvelilac,dashed,mark=*},
  sblong/.style={curvepurple,solid,mark=square*},
  bp/.style={curvered,dashdotted,mark=+},
  bpmax/.style={curvegreen,solid,mark=triangle*,mark options={rotate=180}},
  sclsmall/.style={black!55,dashdotted,mark=+},
  scllarge/.style={black!85,dashed,mark=x},
  ml/.style={black!85,dashed,mark=diamond,mark options={solid,fill=white}},
}
\begin{tikzpicture}
\begin{groupplot}[
  group style={group size=2 by 1,horizontal sep=4.4cm},
  width=3.5cm,height=3.5cm,scale only axis,
  ymode=log,log basis y=10,xmin=0.95,xmax=3.55,
  xtick={1,1.5,2,2.5,3,3.5},xlabel={$E_b/N_0$ (dB)},
  label style={font=\scriptsize},tick label style={font=\scriptsize},
  title style={font=\small},
  axis line style={black!65},tick style={black!65},
  ymajorgrids,xmajorgrids,grid style={gray!20},minor y tick num=0,
  every axis plot/.append style={line width=0.65pt,mark size=1.2pt},
  legend style={draw=none,fill=none,font=\scriptsize,
    /tikz/every even column/.append style={column sep=3pt}},
  legend columns=1,
]
\input{figures/polar12864-decoder-comparison}
\input{figures/ccsds-decoder-comparison}
\end{groupplot}
\path (group c1r1.east) -- (group c2r1.west) coordinate[midway] (legendcenter);
\node[anchor=center] at ([xshift=-4pt]legendcenter) {\pgfplotslegendfromname{polar-ccsds-legend}};
\end{tikzpicture}
\endgroup
\vspace{-4.5mm}
\caption{\henrytask{1,11}{BLER over BPSK-AWGN. \revision2{\mbsresult{2048}} on the CCSDS and Polar codes illustrated in Figure~\ref{fig:polar-masking-motivation}.}}
\label{fig:polar-ccsds-bler}
\label{fig:polar12864-decoder-comparison}
\label{fig:ccsds-step-comparison}
\end{figure}

\paragraph{Codes, channels, and model.}
We consider LDPC~\citep{gallagerLowDensityParityCheck1962}, Polar~\citep{arikanChannelPolarization2009}, and Bose--Chaudhuri--Hocquenghem (BCH) codes~\citep{hocquenghemCodesCorrecteurs1959,boseClassErrorCorrecting1960}, including LDPC codes from MacKay\textquotesingle s collection~\citep{mackaySparseGraphCodeDatabase} and the Consultative Committee for Space Data Systems (CCSDS) short-block family~\citep{ccsdsShortBlockLDPC2015}.
For the physical channel, we consider BPSK over AWGN.

\ifdefined\preprintupdatedresults
The BPSK-AWGN comparison in Table~\ref{tab:short-code-ber} considers $E_b/N_0\in\{4,5,6\}$ dB.
\else
Our BPSK-AWGN evaluations in Table~\ref{tab:short-code-ber} cover linear block codes at $E_b/N_0\in\{4,5,6\}$ dB.
\fi
The Polar$(128,64)$ and CCSDS$(128,64)$ curves in Figure~\ref{fig:polar-ccsds-bler}(a)--(b) additionally cover $E_b/N_0=1$--$3.5$ dB in $0.5$ dB increments.
We use a CrossMPT~\citep{parkCrossMPTCrossAttentionMessagePassing2025} architecture with 6 attention layers, width 128, and 8 attention heads to model $Q$; additional implementation details are in Appendix~\ref{app:posterior-estimator-implementation}.

\paragraph{Training and evaluation.}
We train with \eqref{eq:posterior-marginal-training-loss} for three million iterations on random codewords, with $E_b/N_0$ sampled uniformly from $\{2,3,4,5,6,7\}$ dB, and evaluate exponential-moving-average (EMA) weights. Unless otherwise stated, diffusion uses 16 confidence-balanced reverse steps \revision2{as defined in Section~\ref{sec:cond-diffusion}}. Appendix~\ref{app:training-simulation-details} gives optimizer and EMA settings, simulation budgets, and stopping criteria.
\ifdefined\preprintupdatedresults\else
The paired campaign evaluations target 500 \revision2{MBS} AWGN frame errors and stop at their configured frame cap; the revised cap is $10^8$ frames, with previously collected \mbox{counters retained.}
\fi

\paragraph{\revision2{MBS}.}
\ifdefined\preprintupdatedresults
For Table~\ref{tab:short-code-ber}, \revision2{MBS} uses maximum \revision2{beam size} $L=\revision2{64}$, initial branching depth $d_0=5$, \revision2{at most $\revision2{K_B}=16$ branching rounds, and} $\revision2{K_C}=16$ \revision2{confidence-balanced} diffusion steps per completion attempt. \revision2{Each candidate branches on its least-confident unresolved bit, and pruning to the beam budget precedes completion. We use} minimum-distance selection~\eqref{eq:maxwell-minimum-distance-decision}.
\else
The decoder in Section~\ref{sec:maxwell} uses beam width $L=32$, initial branching depth $d_0=5$, at most $\revision2{K_B}=16$ branching rounds, and $\revision2{K_C}=8$ balanced diffusion steps per completion attempt.
\fi

\paragraph{Baselines.}
For theoretical decoding baselines, we use belief propagation (BP)~\citep{kschischangFactorGraphsSumProduct2001}, successive cancellation (SC)~\citep{arikanChannelPolarization2009}, and successive-cancellation list (SCL) decoding~\citep{talListDecodingPolar2015}.
Table~\ref{tab:short-code-ber} \revision2{compares BP,} CrossMPT~\citep{parkCrossMPTCrossAttentionMessagePassing2025}, DDECC~\citep{choukrounDenoisingDiffusionError2023}, SB-ECC~\citep{helvitsScoreBasedError2026}\revision2{, our conditional diffusion decoder, and \revision2{MBS} with beam size 64}.

\subsection{Results}

\ifdefined\preprintupdatedresults
\revision2{MBS} \revision2{with $L=64$} achieves the \revision2{highest} reported \revision2{$-\ln(\mathrm{BER})$} among \revision2{the six} methods in all 42 code--SNR comparisons in Table~\ref{tab:short-code-ber}. \revision2{MBS improves on conditional diffusion throughout. SB-ECC has lower reported BER than our conditional diffusion in 39 of the 42 comparisons.} \revision2{At} 6 dB, CCSDS$(128,64)$ \revision2{has} no observed errors over $10^9$ frames. \revision2{Its dagger value $22.94$ represents the lower bound obtained by assuming} one frame error with $d_{\min}\revision2{=14}$ bit errors\revision2{. Evaluation details are given in Appendix~\ref{app:training-simulation-details}}.
\else
Table~\ref{tab:short-code-ber} summarizes BER performance over BPSK-AWGN. SB-ECC has lower reported BER than our conditional diffusion decoder in 39 of the 42 code--SNR comparisons; the exceptions are Polar$(64,32)$ and Polar$(128,96)$ at 5 dB, and CCSDS$(128,64)$ at 6 dB; for the latter, the campaign diffusion estimate is based on one frame error. One possible explanation is that DDECC and SB-ECC are formulated as reversing the channel noise, which provides an inductive bias suited to AWGN. Augmenting our decoder with \revision2{MBS} yields the highest finite negative-log-BER point estimate in 37 of the 41 cells with a finite \revision2{MBS} estimate. The remaining CCSDS$(128,64)$ cell at 6 dB has zero observed \revision2{MBS} errors and no finite estimate; frame-cap results with few errors remain statistically uncertain.
\fi

\par
Figure~\ref{fig:polar-ccsds-bler}(a)--(b) compares block error rates (BLER) on Polar$(128,64)$ and CCSDS$(128,64)$, the same codes considered in the conservation-law example of Figure~\ref{fig:polar-masking-motivation}. We train SB-ECC for three million steps and evaluate it with up to 100 denoising steps, while our plain conditional-diffusion decoder uses up to 128 steps \revision2{(any-order AR)}. These larger evaluation budgets aim to maximize the performance of purely diffusion-based decoding. Our conditional-diffusion decoder fixes one bit at a time in confidence order, analogous to autoregressive decoding along a chain-rule factorization. Both methods retain parity-based early exit. We further test \revision2{MBS} with BP replacing our learned model to estimate the posterior marginals. Additional details are given in Appendix~\ref{app:polar-ccsds-curves}. \revision2{Appendix~\ref{app:list-cap-performance} examines the effect of the maximum beam size $L$ on MBS decoding performance and reports mean beam usage and its empirical percentiles for CCSDS$(128,64)$ at 2.5 and 4\,dB.}

The advantage of \revision2{MBS} over the neural diffusion baselines grows at higher SNR. On CCSDS, using BP estimates within \revision2{MBS} also outperforms both plain conditional diffusion and SB-ECC, approaching the performance of learned \revision2{MBS}. On Polar, however, the parity-check graph has many short cycles and highly uneven node degrees, which may contribute to BP's poor estimates. \revision2{MBS} improves Polar BP only modestly and does not overcome this limitation. Learned \revision2{MBS} still has a gap to the CCSDS ML reference~\citep{helmlingChannelCodes2025} and the Polar SCL references with list sizes 32 and 512.
\par

\subsection{Decoding latency and stopping-time distribution}
\label{sec:decoding-latency}

\begin{figure}[!t]
\centering
\begin{tikzpicture}
\begin{groupplot}[
  group style={group size=4 by 1,horizontal sep=0.45cm},
  width=2.7cm,height=1.2cm,scale only axis,
  clip=false,
  xmin=0,xmax=100,
  ymode=log,log basis y=10,ymin=0.00007,ymax=1.3,
  xtick={0,50,100},xticklabels={0,50,100},
  ytick={0.0001,0.01,1},yticklabels={$10^{-4}$,$10^{-2}$,$1$},
  minor tick num=0,tick pos=left,tick align=outside,
  tick label style={font=\scriptsize},label style={font=\scriptsize},
  axis x line*=bottom,axis y line*=left,
  axis line style={black!45,line width=0.4pt},tick style={black!45},
  ymajorgrids,grid style={black!10,line width=0.3pt},
  axis background/.style={fill=black!1},
]
\input{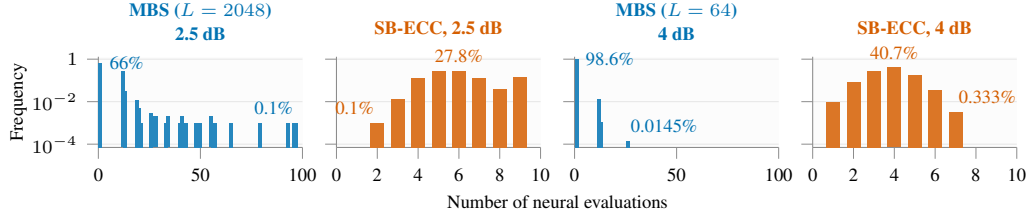}
\end{groupplot}

\node[rotate=90,font=\scriptsize,inner sep=0pt] at ([xshift=-30pt]group c1r1.west)
  {Frequency};
\path (group c1r1.south west) -- (group c4r1.south east) coordinate[midway] (histxlabel);
\node[font=\scriptsize,anchor=north,inner sep=0pt] at ([yshift=-18pt]histxlabel)
  {Number of neural evaluations};
\end{tikzpicture}
\caption{\henrytask{1,4}{Distribution of the number of neural function evaluations (NFE) for \revision2{MBS\ifdefined\arxivlayout\ ($L=2048$ at 2.5 dB; $L=64$ at 4 dB)\fi} and SB-ECC on CCSDS$(128,64)$ at 2.5 dB (left pair) and 4 dB (right pair).}}
\label{fig:ccsds-stopping-time}
\end{figure}

\henrytask{4,10}{For 1,000 CCSDS$(128,64)$ frames at $E_b/N_0=2.5$ dB, we record the number $N$ of sequential neural evaluations performed before decoding terminates. We count each parallel batch of beam candidates as one evaluation, including the initial greedy prediction, branching queries, and all \mbox{completion attempts.}
Based on measurements on an RTX A5000, the estimated mean inference times for \revision2{MBS} and SB-ECC are 47.03 and 40.01 ms at 2.5 dB, and 9.30 and 25.01 ms at 4 dB, respectively. At 4 dB, \revision2{MBS} stops after one neural evaluation on 98.55\% of frames, giving it a lower mean inference time than SB-ECC. This early exit is enabled by direct bitwise posterior predictions, whereas SB-ECC uses conditional noise means for iterative denoising. At 2.5 dB, only 66\% of frames stop after one evaluation; the additional search on the remaining frames raises \revision2{MBS}'s mean inference time above that of SB-ECC. Further details are given in Appendix~\ref{app:complexity-latency}.
}

\paragraph{Additional experiments and analysis.}
Beyond the comparisons above, we also evaluate longer LDPC codes in Appendix~\ref{app:longer-code-results}. Appendix~\ref{app:channel-extensions} studies generalization to Rayleigh fading, and Appendix~\ref{app:complexity-latency} reports computational complexity and measured decoding latency.

\par

\section{Conclusion}

We develop a conditional masked diffusion framework for error-correcting codes in which the received word remains fixed side information. Learned posterior marginals guide the reverse process, motivated by an information-theoretic conservation law. The masked formulation enables a beam search mechanism inspired by the Maxwell decoder of \citet{meassonMaxwellConstructionHidden2008}. Our experiments show that augmenting conditional diffusion with \revision2{MBS} yields competitive decoding performance across the evaluated settings.

Reliable ECC decoding is important for communication and storage, and offers a controlled setting for studying learned inference under exact constraints. We believe that the \revision2{MBS} mechanism may also benefit diffusion-based generation more broadly by retaining alternative partial solutions before committing to uncertain decisions. Related uses of search during generation are already being explored in SOAR~\citep{caoSearchAccelerate2026} and Diffusion Tree Sampling~\citep{jainDiffusionTreeSampling2025}, suggesting a promising direction for extending these ideas beyond ECC.
\par

\section*{AI Use Statement}
We used OpenAI Codex to assist with finding and summarizing related work; evaluating methodological and experimental options; implementing, debugging, and analyzing research code and evaluations; and drafting, organizing, and editing portions of the manuscript.
We treated these outputs as suggestions and reviewed AI-assisted text, code, citations, and results before incorporating them.
The authors take responsibility for all claims, results, and content of this paper.
\par

\ifdefined\arxivlayout
\subsubsection*{Acknowledgments}
The work of H. D. Pfister and Z. Aharoni was supported in part by the National Science Foundation (NSF) under Grant Number 2308445.
\fi

\bibliography{references}
\bibliographystyle{iclr2027_conference}

\ifdefined\arxivlayout\newpage\else\clearpage\fi
\appendix
\section*{Appendix Contents}
\revision2{Appendix~\ref{app:conditional-conservation-laws} proves the conditional entropy and cross-entropy conservation laws for masking.} Appendix~\ref{app:parallel-score-expectation} relates the expected parallel decoding score to conditional dependence and posterior approximation error. Appendix~\ref{app:posterior-estimator-implementation} describes the neural posterior estimator and its conditioning on physical and auxiliary observations.

Appendix~\ref{app:evaluation-details} documents evaluation protocols, sample sizes, and reference results. Appendix~\ref{app:additional-decoding-results} presents longer-code results. Appendix~\ref{app:channel-extensions} studies generalization to Rayleigh fading. Appendix~\ref{app:list-cap-performance} examines the effect of the maximum \revision2{beam} size in MBS. Appendix~\ref{app:complexity-latency} analyzes computational complexity and measured decoding latency.

\par
\medskip
\begingroup
\setlength{\parskip}{2pt}
\newcommand{\appendixcontentsentry}[1]{%
  \noindent\makebox[1.5em][l]{\ref{#1}}\nameref{#1}\nobreak\dotfill\pageref{#1}\par}
\appendixcontentsentry{app:conditional-conservation-laws}
\appendixcontentsentry{app:parallel-score-expectation}
\appendixcontentsentry{app:posterior-estimator-implementation}
\appendixcontentsentry{app:evaluation-details}
\appendixcontentsentry{app:additional-decoding-results}
\appendixcontentsentry{app:channel-extensions}
\appendixcontentsentry{app:list-cap-performance}
\appendixcontentsentry{app:extended-related-work}
\endgroup

\section{\texorpdfstring{\revision2{Conditional Conservation Laws for Masking}}{Conditional Conservation Laws for Masking}}
\label{app:conditional-conservation-laws}
\label{sec:conservation-laws}
\revision2{
We prove the conditional masking identities in Section~\ref{sec:masked-conservation-law}, adapting \citet[Appendices~A.4--A.5]{aharoniConservationLawsDiffusion2026}. Let $P_{X\mid R}$ and $Q_{X\mid R}$ be the true and model posteriors induced by the shared physical channel. Assume $P_{X\mid R}\ll Q_{X\mid R}$ almost surely and finite conditional cross-entropy. Under both laws, each coordinate is independently masked with probability $t_i$, independently of $R$ given $X$. All expectations below use the true law.
}

\paragraph{\texorpdfstring{\revision2{Proof.}}{Proof.}}
\label{app:conditional-masking-proof}
\revision2{
For coordinatewise masking probabilities $\mathbf t$, put
\[
 F(\mathbf t)=-\mathbb E\log Q_{X\mid Y(\mathbf t),R}(X\mid Y(\mathbf t),R).
\]
Fix $\mathbf t_{-i}$ and abbreviate $Z_i=(Y_{-i}(\mathbf t_{-i}),R)$.
Conditional independence under both laws gives the cross-entropy chain rule
\begin{align*}
 F(\mathbf t)
 &=-\mathbb E\log Q_{X_i\mid Y_i(t_i),Z_i}(X_i\mid Y_i(t_i),Z_i)\\
 &\quad-\mathbb E\log Q_{X_{-i}\mid X_i,Z_i}(X_{-i}\mid X_i,Z_i).
\end{align*}
The second term is independent of $t_i$. For the first term,
\[
 Q_{X_i\mid Y_i(t_i),Z_i}(X_i\mid Y_i(t_i),Z_i)
 =\begin{cases}
 1, & Y_i(t_i)=X_i,\\
 Q_{X_i\mid Z_i}(X_i\mid Z_i), & Y_i(t_i)=\ast.
 \end{cases}
\]
Since the mask is independent of $(X,Z_i)$,
\[
 -\mathbb E\log Q_{X_i\mid Y_i(t_i),Z_i}(X_i\mid Y_i(t_i),Z_i)
 =(1-t_i)\underbrace{(-\log 1)}_{0}
 +t_i\mathbb E[-\log Q_{X_i\mid Z_i}(X_i\mid Z_i)].
\]
Hence
\begin{equation}
 \frac{\partial F}{\partial t_i}(\mathbf t)
 =-\mathbb E\log Q_{X_i\mid Z_i}(X_i\mid Z_i).
 \label{eq:conditional-masking-local-ce}
\end{equation}
Along $\mathbf t=(t,\ldots,t)$, the path derivative is the sum of the coordinate derivatives in~\eqref{eq:conditional-masking-local-ce}, giving $\frac{d}{dt}F(t,\ldots,t)=\mathcal C_{P,Q}(t)$. The fundamental theorem of calculus then yields~\eqref{eq:conditional-mismatched-masked-conservation}, with the endpoint conditions satisfied because $Y(0)=X$ and $Y(1)=\ast^n$:
\[
 F(0,\ldots,0)=0,\qquad
 F(1,\ldots,1)=\operatorname{CE}(P_{X\mid R},Q_{X\mid R}).
\]
}

\revision2{Setting $Q=P$ gives the entropy derivative~\eqref{eq:conditional-masked-gexit} and, with the same endpoints, the conservation law~\eqref{eq:conditional-masked-conservation}.
}

\FloatBarrier
\section{Parallel Sampling, Total Correlation, and Path Scores}
\label{app:parallel-score-expectation}

\henrytask{7}{Independent sampling replaces the joint posterior of the selected bits by a product of marginals.
Total correlation (TC)\revision2{, introduced by \citet{watanabeInformationTheoretical1960},} measures the resulting expected log-score penalty.}

\revision2{TC and dual TC have been used to analyze parallel sampling schedules for masked diffusion~\citep{chenOptimalInferenceSchedules2025}. \citet{wenConditionalTotalCorrelation2026} characterize the forward KL divergence of adaptive parallel sampling as accumulated conditional TC. Here we apply this dependence-penalty perspective to the decoder's path score, conditioned on the physical observation $R$.}

\paragraph{Penalty for a parallel step.}
Fix $R=r$ and the decoder state $\widetilde y^{(k)}$.
Select $\mathcal G=\mathcal G_k$ before assigning its bit values, and write
\[
  \pi(x_{\mathcal G})
  =P_{X_{\mathcal G}\mid \revision2{Y(t^{(k)}),R}}(x_{\mathcal G}\mid \revision2{\widetilde y^{(k)},r}),
  \qquad
  \pi_i(x_i)
  =P_{X_i\mid \revision2{Y(t^{(k)}),R}}(x_i\mid \revision2{\widetilde y^{(k)},r}).
\]
The expected difference between the joint log probability and the sum of marginal log \mbox{probabilities is}
\begin{align}
  \operatorname{TC}(X_{\mathcal G}\mid \revision2{\widetilde y^{(k)},r})
  &:=D_{\mathrm{KL}}\!\left(\pi\,\middle\|\,\prod_{i\in\mathcal G}\pi_i\right)\notag\\
  &=\mathbb E_\pi\!\left[\log\pi(X_{\mathcal G})
    -\sum_{i\in\mathcal G}\log\pi_i(X_i)\right]\geq0.
  \label{eq:parallel-score-total-correlation}
\end{align}
TC vanishes for conditionally independent selected bits, and automatically for a singleton.

\par
\paragraph{Accumulated path score.}
\henrytask{7}{To isolate the effect of parallel reveals, we analyze the path score $s(x)$ in \eqref{eq:maxwell-score-decision} using exact posterior marginals.
Each group is selected using $r$ and the previously revealed bits.
Predicting and revealing one bit at a time takes $K=n$ steps for a length-$n$ codeword.
Updating the exact posterior after each bit, the probability chain rule gives $s(x)=\log P_{X\mid R}(x\mid r)$.
When $K<n$, some steps reveal several bits in parallel; summing their local TC penalties gives
\begin{align}
  \mathbb E[-s(X)\mid R=r]
  ={}&H(X\mid R=r)\notag\\
    &+\sum_{k=1}^{K}\mathbb E[\operatorname{TC}(X_{\mathcal G_k}\mid \revision2{\widetilde y^{(k)},r})\mid R=r].
  \label{eq:parallel-path-tc-penalty}
\end{align}
Thus the expected negative path score upper-bounds posterior entropy.}

\paragraph{Learned marginals.}
Let $q_i$ denote $Q_{X_i\mid \revision2{Y(t^{(k)}),R}}(\cdot\mid \revision2{\widetilde y^{(k)},r})$.
For finite log loss and model support covering the true support,
\begin{equation}
  \mathbb E_\pi\!\left[\log\pi(X_{\mathcal G})-\sum_{i\in\mathcal G}\log q_i(X_i)\right]
  =\operatorname{TC}(X_{\mathcal G}\mid \revision2{\widetilde y^{(k)},r})
   +\sum_{i\in\mathcal G}D_{\mathrm{KL}}(\pi_i\|q_i).
  \label{eq:learned-parallel-score-gap}
\end{equation}
Marginal prediction error therefore adds to the dependence penalty.

\par

\FloatBarrier
\section{Posterior-Estimator Implementation}
\label{app:posterior-estimator-implementation}

We implement the posterior estimator of Section~\ref{sec:decoder-architecture} with a CrossMPT backbone \citep{parkCrossMPTCrossAttentionMessagePassing2025}.
Code-bit and parity-check tokens use alternating $H$- and $H^{\mathsf T}$-masked cross-attention.

At decoding step $k$, an auxiliary encoder represents the current state and mask fraction $(\widetilde y^{(k)},t^{(k)})$.
A separate physical encoder represents the fixed observation $r$ and any available physical-channel parameters.
Both encoders condition every CrossMPT layer through adaptive layer normalization: their scale contributions are added, and their shift contributions are added.
The alternating cross-attention blocks remain unchanged.

For each masked bit $i$, the output head predicts logits for the extrinsic posterior
\begin{equation}
  Q_{X_i\mid \revision2{Y_{-i}(t^{(k)}),R}}
    \!\left(x_i\mid \revision2{\widetilde y_{-i}^{(k)},r}\right),
  \qquad x_i\in\{0,1\}.
  \label{eq:decoder-step-extrinsic-marginals}
\end{equation}
The known local auxiliary-channel likelihood is applied outside the network using \eqref{eq:masked-bitwise-posterior}.
Thus, $Q_{X_i\mid \revision2{Y(t^{(k)}),R}}(x_i\mid \revision2{\widetilde y^{(k)},r})$ equals the predicted extrinsic posterior at masked positions and $\mathbf 1\{x_i=\widetilde y_i^{(k)}\}$ at revealed positions.
These marginals drive both the diffusion and \revision2{MBS}~\mbox{decoders.}
\par

\FloatBarrier
\section{Evaluation Protocols and Reference Results}
\label{app:evaluation-details}

This appendix describes the training procedure, simulation protocol, reference sources, decoder configurations, sample sizes, and stopping criteria used in the experimental comparisons. Comparisons with published methods use independently generated samples and may differ in training objective, inference procedure, and evaluation budget.

\subsection{Baselines evaluations}
The BP, CrossMPT, DDECC, and SB-ECC BER values in Table~\ref{tab:short-code-ber} are transcribed from \citet[Table~1]{helvitsScoreBasedError2026}. BP uses 50 decoding iterations, and DDECC and SB-ECC share the CrossMPT backbone in that comparison.

For LDPC$(49,24)$, DDECC reports at least $10^5$ frames and at least 500 erroneous frames per SNR; SB-ECC stops after 500 frame errors or $10^9$ evaluated LDPC frames; CrossMPT reports at least 500 erroneous frames without specifying the total frame count. At 4 dB, \citet[Appendix Table~3]{parkCrossMPTCrossAttentionMessagePassing2025} report $-\ln(\mathrm{BLER})=4.47$ for CrossMPT, where BLER denotes the frame-error rate.
For the longer-code comparison in Table~\ref{tab:long-code-ber}, the SB-ECC values for LDPC$(204,102)$ and LDPC$(529,440)$ are transcribed from \citet[Table~6]{helvitsScoreBasedError2026}.

\FloatBarrier

\subsection{Training and simulation details}
\label{app:training-simulation-details}

\paragraph{AWGN training.}
We train a separate posterior estimator for each code. Training samples are generated online from random codewords, with $E_b/N_0$ drawn uniformly from the discrete grid $\{2,3,4,5,6,7\}$~dB. For every sample, the codeword is mapped to BPSK and passed through AWGN to obtain the fixed physical observation. We then draw $t\sim\operatorname{Unif}(0,1)$ independently for each sequence, mask coordinates independently with probability $t$, and condition on masking at least one coordinate. The model minimizes the masked-bit negative log-likelihood in~\eqref{eq:posterior-marginal-training-loss} for three million iterations. We use batches of 256 codewords and AdamW with zero weight decay; the learning rate is cosine-annealed from $5\times10^{-4}$ to $5\times10^{-6}$. We maintain both raw and exponential-moving-average (EMA) model parameters during training and use the EMA weights for the reported AWGN diffusion and \revision2{MBS} evaluations. \ifdefined\preprintupdatedresultsThe EMA decay factor is $0.9999$.\fi{} The posterior-estimator architecture is specified in Appendix~\ref{app:posterior-estimator-implementation}. Unless otherwise stated, conditional diffusion uses 16 confidence-balanced reverse steps with equal-sized reveal groups, up to integer rounding, fixing the most confident masked \mbox{coordinates first.}
\paragraph{Monte Carlo simulation.}
For a rate-$R_c=k/n$ code at a specified $E_b/N_0$, each frame independently samples a random codeword and Gaussian noise according to
\begin{equation}
  r_i=1-2x_i+z_i,
  \qquad z_i\overset{\mathrm{i.i.d.}}{\sim}\mathcal N(0,\sigma^2),
  \qquad \sigma^2=\frac{1}{2R_c10^{(E_b/N_0)_{\mathrm{dB}}/10}}.
\end{equation}
\ifdefined\preprintupdatedresults
\paragraph{Diffusion evaluation.}
\revision2{For Table~\ref{tab:short-code-ber}, conditional diffusion uses EMA weights and up to 16 confidence-balanced reverse steps with parity-based early exit. The Polar evaluations first attempt a one-step greedy completion; frames with nonzero syndrome restart diffusion from the original channel observation. Polar$(64,32)$ uses the three-million-update no-bias checkpoint; Polar$(128,86)$ and Polar$(128,96)$ use the 2.7-million- and 2.8-million-update EMA checkpoints, selected by validation FER at 4\,dB, with BER breaking ties. The standard stopping rule is 500 frame errors or $10^8$ frames ($10^9$ for LDPC codes), whichever occurs first; the paired Polar evaluations use their saved evaluation populations, whose sizes and error counts are recorded in the accompanying artifacts. BER and FER include all evaluated frames, including incorrect parity-valid greedy outputs.}
\par
\paragraph{Maxwell beam-search evaluation.}
\revision2{For Table~\ref{tab:short-code-ber}, MBS uses the corresponding saved-bank checkpoints, maximum beam size $L=64$, initial depth $d_0=5$, at most $\revision2{K_B}=16$ branching rounds, and $\revision2{K_C}=16$ confidence-balanced completion steps. Each candidate branches on its least-confident unresolved bit; pruning precedes completion, and no confident-variable decimation is used. We replay the saved greedy-invalid frames while retaining the original greedy decisions and total-frame denominators, including incorrect greedy outputs. Each selected bank is processed in full, without a new 500-error stopping criterion. Per-cell sample sizes, observed errors, checkpoint hashes, and bank provenance are provided in the beam-size comparison artifact. The extended 6\,dB populations for LDPC$(121,60)$, LDPC$(121,70)$, LDPC$(121,80)$, and CCSDS$(128,64)$ each contain $10^9$ frames. For CCSDS, zero observed errors are displayed as the resolution marker $-\ln[d_{\min}/(nN)]=22.94$, with $d_{\min}=14$, rather than a confidence bound.}
\else
Unless stated otherwise, conditional diffusion uses $K=16$ confidence-balanced reverse steps. \revision2{MBS} uses beam width $L=16$, fifteen single-bit branches at the least-confident unresolved positions, parity peeling, and one final parallel completion. The model-path and minimum-distance decision rules are evaluated on the same frames and candidate \revision2{beams}. Diffusion evaluations stop after 500 frame errors or $10^8$ frames.
\fi

\paragraph{Compute.}
\ifdefined\preprintupdatedresults
\revision2{Training and the beam-64 saved-frame replay used NVIDIA H200 GPUs. Original evaluations used NVIDIA RTX A5000 GPUs, with the Polar evaluations using RTX 2080 Ti or H200 GPUs.}
\else
The base models were trained on NVIDIA H200 graphics processing units (GPUs). The short-code simulations (codes with $n\le 128$) were evaluated on NVIDIA RTX 2080 Ti GPUs, except for the updated Polar$(64,32)$ evaluation on H200, whereas the longer-code experiments in Appendix~\ref{app:additional-decoding-results} used H200 GPUs. These hardware statements describe where our runs were executed; the comparisons with published baselines are not \mbox{compute matched.}
\fi

\subsection{Polar and CCSDS decoder-comparison curves}
\label{app:polar-ccsds-curves}

Figure~\ref{fig:polar-ccsds-bler}(a)--(b) uses BPSK-AWGN at rate $1/2$, with $E_b/N_0=1,1.5,\ldots,3.5$ dB. Our plain-diffusion and \revision2{MBS} variants share the respective code's three-million-update EMA checkpoint. SB-ECC uses three-million-update checkpoints for both Polar and CCSDS. SB-ECC uses a solver setting of 101 steps, corresponding to at most 100 Euler updates, with syndrome-based early exit. Our conditional-diffusion decoder uses up to 128 confidence-ordered passes with parity-based early exit. For both methods, we increased the number of diffusion steps until decoding performance saturated.

We run \revision2{MBS} with either our neural model or BP, each providing estimates of the bitwise posterior marginals conditioned on the received word and the candidate's revealed bits. In the BP variant, each neural-model call is replaced by 50 iterations of sum-product BP. Both variants use $d_0=5$, $\revision2{K_B}=16$, $L=2048$, and $\revision2{K_C}=8$, with parity propagation and balanced completion steps. We use the minimum-distance decision rule in \eqref{eq:maxwell-minimum-distance-decision}.

\henrytask{12}{For each BP call, we initialize fresh messages from the AWGN channel LLRs $2r_i/\sigma^2$ and run 50 flooding sum-product iterations~\citep{kschischangFactorGraphsSumProduct2001}, updating all check-to-variable messages and then all variable-to-check messages. Check-node updates use the log-domain sum-product rule, without a min-sum approximation~\citep{fossorier1999reduced}. Revealed bits are clamped to their assigned values and send signed infinite outgoing LLRs. Computation uses float64 arithmetic, with finite LLRs clipped to $[-50,50]$. We apply no message scaling, offset correction, or damping (scale 1, offset 0, damping 0). Polar BP operates on the common standard-form parity-check matrix, not the polar-transform factor graph.}

The SCL baselines~\citep{talListDecodingPolar2015} use list sizes 32 and 512, with frozen bits matching the parity-check matrix used in our simulations. Every Polar point stops at exactly the 100th frame error. The external CCSDS ML reference has 50 errors per point.


\FloatBarrier

\subsection{Polar$(64,32)$ BER comparison}
\label{app:polar64-ber-sources}

\begin{table}[htbp]
\centering
\caption{\henrytask{1}{Coded-bit BER for Polar$(64,32)$ at 4 dB, as shown in Figure~\ref{fig:polar-latency-fer-placeholder}. Values are $-\ln(\mathrm{BER})$; higher is better.}}
\label{tab:polar64-ber-summary}
\small
\begin{tabular}{lr}
\toprule
Decoder & $-\ln(\mathrm{BER})$ \\
\midrule
BP-5 & 3.53 \\
BP-10 & 3.80 \\
BP-20 & 4.00 \\
BP-50 & 4.29 \\
SCL-1 & 3.81 \\
SCL-4 & 5.88 \\
SCL-8 & 6.87 \\
SCL-16 & 7.56 \\
SCL-32 & 7.83 \\
SCL-64 & 7.92 \\
ECCT & 6.99 \\
CrossMPT & 7.50 \\
DDECC & 7.03 \\
SB-ECC & 7.77 \\
Diffusion & 7.70 \\
\revision2{MBS ($L=64$)} & 7.92 \\
OSD-3 estimate & 7.98 \\
\bottomrule
\end{tabular}
\end{table}

Table~\ref{tab:polar64-ber-summary} supplies the coded-bit BER values in Figure~\ref{fig:polar-latency-fer-placeholder}. ECCT and CrossMPT values come from Table~4 of \citet{parkCrossMPTCrossAttentionMessagePassing2025}; DDECC, SB-ECC, and conditional diffusion follow Table~\ref{tab:short-code-ber}.

\paragraph{BP and Polar SCL baselines.}
We evaluate sum-product BP on the Tanner graph of the code's parity-check matrix, using 5, 10, 20, or 50 iterations \citep{kschischangFactorGraphsSumProduct2001}. CRC-free SCL uses list sizes $L\in\{1,4,8,16,32,64\}$ \citep{talListDecodingPolar2015}. It re-encodes the decoded information vector to measure BER over all 64 codeword positions. Both baselines use the same Polar$(64,32)$ code.

\paragraph{\revision2{MBS} and OSD reference.}
The updated 4 dB diffusion and \revision2{MBS} estimates use the same 177,664 frames, with 5,134 and \revision2{4,120} bit errors, respectively. \revision2{The \revision2{MBS} result in Figure~\ref{fig:polar-latency-fer-placeholder} uses least-confident branching with beam size 64 and has 502 frame errors, giving $-\ln(\mathrm{BER})=7.92292$, which rounds to 7.92.} The decoder settings and \revision2{saved-frame replay protocol} are given above. OSD-3 perturbs the hard decisions on a most-reliable information set in up to three positions, re-encodes each candidate, and selects the codeword with minimum Euclidean distance to the received word \citep{fossorierSoftDecisionDecodingLinear1995}. Its BER estimate uses 8,776 coded-bit errors over 400,000 frames at 4 dB.

\FloatBarrier


\FloatBarrier
\section{Longer-Code Results}
\label{app:additional-decoding-results}
\label{app:longer-code-results}

\par
This appendix evaluates decoding performance for two longer LDPC codes over BPSK--AWGN at 4 and 5~dB.
The models use the training and simulation protocol in Appendix~\ref{app:training-simulation-details}.
We report $-\ln(\mathrm{BER})$.
\revision2{MBS} yields the strongest point estimates across the reported SNRs. Diffusion is slightly above SB-ECC for LDPC$(204,102)$ at 4 dB and below it in the other \mbox{three comparisons.} We did not evaluate at 6~dB, as the low error rates already observed at 5~dB would require substantially longer Monte Carlo simulations to obtain reliable estimates at the higher SNR.

\begin{table}[H]
  \centering
  \begin{minipage}{0.82\textwidth}
  \centering
\caption{Negative log BER for long-block LDPC codes over BPSK--AWGN at 4 and 5~dB. SB-ECC results are reproduced from \citet[Table~6]{helvitsScoreBasedError2026}. Boldface and underlining mark the best and second-best distinct values.}
  \label{tab:long-code-ber}
\footnotesize
  \setlength{\tabcolsep}{3pt}
  \begin{tabular*}{\linewidth}{@{\extracolsep{\fill}}l*{3}{rr}@{}}
    \toprule
    Code & \multicolumn{2}{c}{SB-ECC} & \multicolumn{2}{c}{Diffusion} & \multicolumn{2}{c}{\revision2{\mbsresult{32}}} \\
    \cmidrule(lr){2-3}\cmidrule(lr){4-5}\cmidrule(lr){6-7}
    $E_b/N_0$ (dB) & 4 & 5 & 4 & 5 & 4 & 5 \\
    \midrule
    LDPC$(204,102)$ & 11.20 & \underline{15.78} & \underline{11.27} & 15.29 & \textbf{12.09} & \textbf{17.82} \\
    LDPC$(529,440)$ & \underline{7.93} & \underline{15.23} & 7.88 & 13.64 & \textbf{8.69} & \textbf{17.77} \\
    \bottomrule
  \end{tabular*}
  \end{minipage}
\end{table}

\FloatBarrier

\FloatBarrier
\section{Generalization to Rayleigh Fading}
\label{app:channel-extensions}

\subsection{Generalization to Rayleigh fading}
\label{app:rayleigh-fading}
Following the channel setting in \citet[Section~7.4 and Table~7]{helvitsScoreBasedError2026}, we consider
\begin{equation}
  R_i=h_i(1-2X_i)+Z_i,
  \qquad h_i\overset{\mathrm{i.i.d.}}{\sim}\operatorname{Rayleigh}(1),
  \qquad Z_i\overset{\mathrm{i.i.d.}}{\sim}\mathcal N(0,\sigma^2).
\end{equation}
The fading amplitudes are independent across coordinates, so this is a memoryless fading channel. We use $\sigma^2=[2(k/n)10^{(E_b/N_0)_{\mathrm{dB}}/10}]^{-1}$; the scale-one Rayleigh distribution has $\mathbb E[h_i^2]=2$. The receiver is given the channel observations without the fading amplitudes.

\par
We evaluate AWGN-trained three-million-update EMA checkpoints without fine-tuning, using 16 confidence-balanced diffusion steps and \revision2{MBS} with maximum beam width $L=2048$. The stopping target is 100 \revision2{MBS} frame errors or $10^8$ frames. At CCSDS$(128,64)$ and 6 dB, search finished at the $10^8$-frame cap with 1,535 bit errors and 61 frame errors, below the 100-error target.

\begin{table}[!htbp]
\centering
\caption{Rayleigh-fading results for AWGN-trained decoders. Entries are $-\ln(\mathrm{BER})$; higher is better. Baselines are from \citet[Table~7]{helvitsScoreBasedError2026}. Bold and underline mark the best and second-best distinct values in each row at the displayed precision, with ties marked equally.}
\label{tab:rayleigh-channel-comparison}
\scriptsize
\setlength{\tabcolsep}{3pt}
\begin{tabular}{@{}lrrrrrrr@{}}
\toprule
Code & $E_b/N_0$ & \multicolumn{4}{c}{Published baselines} & \multicolumn{2}{c}{Ours} \\
\cmidrule(lr){3-6}\cmidrule(lr){7-8}
& (dB) & BP-50 & BP-200 & CrossMPT & SB-ECC & Diffusion & \revision2{\mbsresult{2048}} \\
& & & & & & 16 steps & Min. distance \\
\midrule
LDPC$(121,70)$ & 4 & 4.06 & 4.26 & 4.25 & \underline{4.62} & 4.32 & \textbf{7.48} \\
LDPC$(121,70)$ & 5 & 5.10 & 5.48 & 5.53 & \underline{6.02} & 5.71 & \textbf{9.58} \\
LDPC$(121,70)$ & 6 & 6.34 & 6.89 & 7.11 & \underline{7.92} & 7.47 & \textbf{12.25} \\
\midrule
CCSDS$(128,64)$ & 4 & 5.25 & 5.59 & 5.25 & \underline{5.77} & 5.53 & \textbf{9.24} \\
CCSDS$(128,64)$ & 5 & 6.78 & 7.37 & 6.94 & \underline{7.70} & 7.46 & \textbf{12.45} \\
CCSDS$(128,64)$ & 6 & 8.72 & 9.64 & 8.92 & \underline{10.08} & 9.81 & \textbf{15.94} \\
\bottomrule
\end{tabular}
\end{table}

Table~\ref{tab:rayleigh-channel-comparison} compares these results with BP-50, BP-200, CrossMPT, and SB-ECC values from \citet[Table~7]{helvitsScoreBasedError2026}. \revision2{MBS} has lower observed BER than diffusion at all six displayed points, and the strongest BER point estimates among the displayed methods. The published baselines use separate evaluations, and the comparison does not control for code construction or SNR normalization across implementations. These values therefore provide a cross-study reference.

\ifdefined\arxivlayout\else\FloatBarrier\fi
\par
\par
\noindent
\ifdefined\arxivlayout\else
\begin{minipage}{\linewidth}
\fi
\revision2{
\section{Beam Size, Computational Complexity, and Latency}
\label{app:list-cap-performance}
\label{app:complexity-latency}
\subsection{Effect of the maximum beam size}
We evaluate MBS with beam sizes $L\in\{64,128,2048\}$ on CCSDS$(128,64)$ and Polar$(128,64)$ over BPSK-AWGN at $E_b/N_0=2.5$ and $4$\,dB, with $d_0=5$, $\revision2{K_B}=16$, and $\revision2{K_C}=8$. We estimate runtime using one initial forward pass, $\revision2{K_C}$ completion calls per attempt, and one branching call between attempts. For a frame with $m$ completion attempts, let $w_j=\min(2^{d_0+j},L)$; its estimated runtime is
\[
\widehat T(m)=t_{\mathrm{fwd}}(1)
+\revision2{K_C}\sum_{j=0}^{m-1}t_{\mathrm{fwd}}(w_j)
+\sum_{j=0}^{m-2}t_{\mathrm{fwd}}(w_j),
\]
with empty sums equal to zero. Here $t_{\mathrm{fwd}}(b)$ is the measured RTX A5000 time for one forward pass with $b$ candidates, linearly interpolated in batch size between measurements at $b\in\{1,64,128,2048\}$. The table gives the mean latency estimates over 100,000 frames at 4\,dB and the full samples at 2.5\,dB. This approximation uses nominal beam growth and $\revision2{K_C}$ calls per probe; it excludes parity-pruning width reductions, early probe termination, and decoder overhead.
\begin{center}
\small
\setlength{\tabcolsep}{3pt}
\begin{tabular}{lrrrr}
\toprule
Code & $E_b/N_0$ (dB) & $L$ & $-\ln(\mathrm{BER})$ & Est. mean (ms) \\
\midrule
CCSDS$(128,64)$ & 2.5 & 64 & 5.24 & 62.2 \\
CCSDS$(128,64)$ & 2.5 & 128 & 5.50 & 75.0 \\
CCSDS$(128,64)$ & 2.5 & 2048 & 6.43 & 140.5 \\
\midrule
Polar$(128,64)$ & 2.5 & 64 & 4.87 & 49.4 \\
Polar$(128,64)$ & 2.5 & 128 & 4.93 & 51.1 \\
Polar$(128,64)$ & 2.5 & 2048 & 5.00 & 55.5 \\
\midrule
CCSDS$(128,64)$ & 4 & 64 & 11.21 & 7.5 \\
CCSDS$(128,64)$ & 4 & 128 & 11.77 & 7.5 \\
CCSDS$(128,64)$ & 4 & 2048 & 13.10 & 7.6 \\
\midrule
Polar$(128,64)$ & 4 & 64 & 9.31 & 10.7 \\
Polar$(128,64)$ & 4 & 128 & 9.37 & 10.7 \\
Polar$(128,64)$ & 4 & 2048 & 9.41 & 10.7 \\
\bottomrule
\end{tabular}
\end{center}
For the BER column, all beam sizes use the same evaluation frames: at 2.5\,dB, 14,720 frames for CCSDS and 11,008 for Polar; at 4\,dB, 7,395,328 frames for CCSDS and 409,856 for Polar.   
}
\ifdefined\arxivlayout\else
\end{minipage}
\fi
\par

\revision2{Table~\ref{tab:ccsds4-mean-beam} reports mean beam usage and its 99th, 99.9th, 99.99th, and 100th percentiles as a function of the maximum beam size for CCSDS$(128,64)$ at $E_b/N_0=2.5$ and $4$\,dB. An initial greedy exit counts as one; P100 denotes the observed maximum.}
\begin{table}[htbp]
\centering
\caption{\revision2{Mean beam usage and empirical percentiles P99, P99.9, P99.99, and P100 for CCSDS$(128,64)$, computed from MBS capped at $L=2048$ over 14,720 frames at 2.5\,dB and 6,215,168 frames at 4\,dB.}}
\label{tab:ccsds4-mean-beam}
\revision2{%
\small\setlength{\tabcolsep}{3pt}
\begin{tabular}{r|rrrrr|rrrrr}
\toprule
 & \multicolumn{5}{c|}{$E_b/N_0=2.5$ dB} & \multicolumn{5}{c}{$E_b/N_0=4$ dB} \\
\cmidrule(lr){2-6}\cmidrule(lr){7-11}
\shortstack{Maximum\\beam size $L$} & Mean & P99 & P99.9 & P99.99 & P100 & Mean & P99 & P99.9 & P99.99 & P100 \\
\midrule
1 & 1.0000 & 1 & 1 & 1 & 1 & 1.0000 & 1 & 1 & 1 & 1 \\
2 & 1.3445 & 2 & 2 & 2 & 2 & 1.0146 & 2 & 2 & 2 & 2 \\
4 & 2.0335 & 4 & 4 & 4 & 4 & 1.0438 & 4 & 4 & 4 & 4 \\
8 & 3.4115 & 8 & 8 & 8 & 8 & 1.1023 & 8 & 8 & 8 & 8 \\
16 & 6.1675 & 16 & 16 & 16 & 16 & 1.2192 & 16 & 16 & 16 & 16 \\
32 & 11.6794 & 32 & 32 & 32 & 32 & 1.4530 & 32 & 32 & 32 & 32 \\
64 & 13.1968 & 64 & 64 & 64 & 64 & 1.4599 & 32 & 32 & 64 & 64 \\
128 & 15.3838 & 128 & 128 & 128 & 128 & 1.4667 & 32 & 32 & 128 & 128 \\
256 & 18.5316 & 256 & 256 & 256 & 256 & 1.4742 & 32 & 32 & 128 & 256 \\
512 & 23.1229 & 512 & 512 & 512 & 512 & 1.4828 & 32 & 32 & 128 & 512 \\
1024 & 29.4881 & 1024 & 1024 & 1024 & 1024 & 1.4922 & 32 & 32 & 128 & 1024 \\
2048 & 37.6968 & 1024 & 2048 & 2048 & 2048 & 1.5014 & 32 & 32 & 128 & 2048 \\
\bottomrule
\end{tabular}}
\end{table}

\ifdefined\arxivlayout\else\FloatBarrier\fi

\subsection{Computational complexity and latency comparison}

\henrytask{4,9}{For neural decoders, $N$ counts sequential forward calls, including all branching and completion queries; a batch of $L$ candidates counts as one call. The number of branching rounds alone does not capture repeated completion attempts, as each completion attempt uses $\revision2{K_C}$ steps.}

\par
\paragraph{Histogram-based time estimates.}
\henrytask{9,10}{For each decoder, we estimate mean latency by averaging over 1,000 simulated frames whose forward-call counts follow the frequencies in the stopping-time histogram, including the full \revision2{MBS} tail. With a per-call latency $t_{\mathrm{fwd}}$, this gives $\widehat T=t_{\mathrm{fwd}}\sum_k k\widehat p_k$. Using forward-pass timings measured on an NVIDIA RTX A5000 GPU, the estimated mean latencies for \revision2{MBS} and SB-ECC are 47.03 and 40.01~ms at 2.5~dB, respectively, and 9.30 and 25.01~ms at 4~dB.}


\begin{table}[htbp]
\centering
\caption{\henrytask{10}{Estimated mean latency for CCSDS$(128,64)$ from the stopping-time histograms in Figure~\ref{fig:ccsds-stopping-time}, using NVIDIA RTX A5000 forward-pass timings.}}
\label{tab:histogram-latency}
\small
\begin{tabular}{lrr}
\toprule
Decoder & 2.5 dB (ms) & 4 dB (ms) \\
\midrule
\revision2{MBS} & 47.03\ifdefined\arxivlayout\ ($L=2048$)\fi & 9.30\ifdefined\arxivlayout\ ($L=64$)\fi \\
SB-ECC & 40.01 & 25.01 \\
\bottomrule
\end{tabular}
\end{table}
\FloatBarrier

\FloatBarrier

\section{Extended Related Work}
\label{app:extended-related-work}

\paragraph{Neural decoders.}
The central line of neural error-correcting code (ECC) decoding consists of iterative, parity-aware methods.
Learned BP decoders explicitly unroll Tanner-graph message passing while training message weights or update rules ~\citep{nachmaniLearningDecodeLinear2016, nachmaniAutoregressiveBeliefPropagation2021}.
The Error Correction Code Transformer (ECCT) and Cross-Attention Message-Passing Transformer (CrossMPT) likewise retain the parity-check structure while replacing BP updates with learned Transformer updates~\citep{choukrounErrorCorrectionCode2022, choukrounFoundationModelError2024, parkCrossMPTCrossAttentionMessagePassing2025}.
Neural polar decoders (NPDs) offer an example of a learned decoder augmented with list decoding, and were proven to be effective in various scenarios \citep{aharoniDatadrivenNeuralPolar2024, hirschStudyNeuralPolar2025}. However, their sequential nature limits parallelism and increases latency, motivating our search for a parallelizable guessing mechanism that can be applied to a broad range of codes.
\paragraph{Decimation and beam-search decoders.}
\henrytask{8}{\citet{buchbergerLearnedDecimationNeural2021} combine least-reliable-bit branching with learned decimation to enhance neural BP for classical LDPC codes. For quantum LDPC codes, \citet{yeBeamSearchDecoder2025} use BP-guided branching and reliability-based beam pruning, while \citet{aydinSoftDecodingQuantum2026} use cumulative BP iterations as a confidence measure for post-selection. Our connection to conditional diffusion conservation laws links the decoder to the Maxwell construction~\citep{meassonMaxwellConstructionHidden2008} and motivates its branching and path pruning: learned posterior marginals guide branching, accumulated reverse-transition log probabilities rank paths, and parity constraints eliminate infeasible candidates.\par}

\paragraph{Diffusion-based ECC decoders.}
Denoising Diffusion Error Correction Codes (DDECC) and the Score-Based Error-Correcting Code decoder (SB-ECC) are recent diffusion-based neural decoders that have achieved competitive performance across a variety of codes~\citep{choukrounDenoisingDiffusionError2023, helvitsScoreBasedError2026}.
Both model decoding as the reversal of channel noise and assume a Gaussian channel.
ECCFM uses soft-syndrome-conditioned consistency training to enable one-step decoding~\citep{leiSyndromeFlowConsistency2026}.
CIDER applies masked diffusion to joint multiuser decoding~\citep{leeStructuredMaskedDiffusion2026}, whereas we study single-codeword decoding with \revision2{MBS}.
Masked diffusion has also been used to infer logical errors from syndrome measurements in quantum LDPC codes~\citep{liuDecodingQuantumDiffusion2025}.
Our conditional formulation separates physical-channel noise from the auxiliary diffusion process, allowing us to choose masking noise and construct a beam decoder.
\par

\paragraph{EXIT, GEXIT, and conservation laws.}
Extrinsic information transfer (EXIT) charts were developed to analyze the behavior of iterative decoders through their extrinsic-information transfer curves, whose area properties connect decoding behavior to code rate~\citep{tenbrinkConvergenceBehaviorIteratively2001a, ashikhminCodeRateArea2002}.
Generalized EXIT (GEXIT) functions and coding conservation laws extend this analysis to more general channel families~\citep{meassonConservationLawsCoding2006, meassonGeneralizedAreaTheorem2009}.
Recently, these conservation laws were extended to diffusion models under posterior mismatch: the data--model cross-entropy, or expected negative log-likelihood, is characterized by an integral along the noise path whose local terms depend on marginal bitwise posteriors~\citep{aharoniConservationLawsDiffusion2026}.
\par

\paragraph{Masked diffusion and search.} Discrete diffusion with an absorbing mask state was studied by \citet{austinStructuredDenoisingDiffusion2021}, and later work showed that its training objective reduces to a weighted masked cross-entropy \citep{sahooSimpleEffectiveMasked2024,shiSimplifiedGeneralizedMasked2024}. Confidence-based parallel masked prediction also appears in Mask-Predict and MaskGIT~\citep{ghazvininejadMaskPredict2019,changMaskGIT2022}. We use the same masking process, but condition it on a fixed physical observation and interpret its loss through coding conservation laws. Search over diffusion trajectories has been explored for generation \citep{jainDiffusionTreeSampling2025,caoSearchAccelerate2026}; our decoder instead exploits hard parity constraints to prune candidates. The \revision2{beam} construction is also related to reliability-based decoders such as Chase decoding \citep{chaseClassAlgorithmsDecoding1972} and ordered-statistics decoding \citep{fossorierSoftDecisionDecodingLinear1995}, which perturb hard decisions according to fixed channel reliabilities.
GRAND and ORBGRAND decode by testing noise patterns ordered by likelihood or channel reliability~\citep{duffyCapacityAchievingGRAND2019,duffyOrderedReliabilityORBGRAND2022}.

\end{document}